\documentstyle[epsfig,twocolumn,subfigure]{mn}

\def\lsim{\mathrel{\rlap{\lower3.5pt\hbox{\hskip0.5pt$\sim$}}
    \raise0.5pt\hbox{$<$}}}
\def\gsim{~\rlap{$>$}{\lower 1.0ex\hbox{$\sim$}}}
\def\ML{\mbox{$M/L$}}
\def\Yst{\mbox{$\Upsilon_{\star}$}}

\title[The MOND fundamental plane]{The MOND Fundamental Plane}
\author[V.F. Cardone et al.]{V.F. Cardone$^{1,2,3}$, G. Angus$^{2}$, A. Diaferio$^{2}$,
C. Tortora$^{4,3}$, R. Molinaro$^{5,6}$\\
$^1$ Dipartimento di Scienze e Tecnologie dell' Ambiente e del Territorio, Universit\`{a} degli Studi del Molise, \\
Contrada Fonte Lappone, 86090\,-\,Pesche (IS), Italy \\
$^2$Dipartimento di Fisica Generale, Universit\`{a} di Torino,
and I.N.F.N. - Sez. di Torino, Via Pietro Giuria 1, 10125 - Torino, Italy \\
$^3$Dipartimento di Scienze Fisiche, Universit\`{a} degli Studi di Napoli  ``Federico II'', \\
Complesso Universitario di Monte S. Angelo, Ed. 6, via Cinthia, 80126 - Napoli, Italy \\
$^4$Institut f\"ur Theoretische Physik, Universit\"at Z\"urich,
Winterthurerstrasse 190, CH\,-\,8057 Z\"urich, Switzerland \\
$^5$I.N.A.F. - Osservatorio Astronomico di Capodimonte, Salita
Moiariello 16, 80131 - Napoli, Italy \\
$^6$Dipartimento di Fisica, Politecnico di Torino, Corso Duca
degli Abruzzi 24, 10125 - Torino, Italy}

\date{Accepted xxx, Received yyy, in original form zzz}

\begin{document}
\maketitle

\begin{abstract}

Modified Newtonian Dynamics (MOND) has been shown to be able to fit spiral galaxy rotation curves as well as giving a theoretical foundation for empirically determined scaling relations, such as the Tully\,-\,Fisher law, without the need for a dark matter halo. As a complementary analysis, one should investigate whether MOND can also reproduce the dynamics of early\,-\,type galaxies (ETGs) without dark matter. As a first step, we here show that MOND can indeed fit the observed central velocity dispersion $\sigma_0$ of a large sample of ETGs assuming a simple MOND interpolating functions and constant anisotropy. We also show that, under some assumptions on the luminosity dependence of the Sersic $n$ parameter and the stellar $M/L$ ratio, MOND predicts a fundamental plane for ETGs\,: a log\,-\,linear relation among the effective radius $R_{eff}$, $\sigma_0$ and the mean effective intensity $\langle I_e \rangle$. However, we predict a tilt between the observed and the MOND fundamental planes.

\end{abstract}


\section{Introduction}

The regularity of their photometric properties and the existence of remarkable scaling relations among their observable quantities may naively suggest that early\,-\,type galaxies (ETGs) are well understood systems. On the contrary, understanding their mass content and density profile is still a hotly debated issues mainly because of both theoretical shortcomings and observational difficulties. Indeed, on the one hand, the lack of a reliable mass tracer makes it difficult to constrain the gravitational potential in the outer (supposedly dark matter dominated) regions, although planetary nebulae surveys (Napolitano et al. 2001, 2002; Romanowsky et al. 2003) are trying to address this problem. On the other hand, the availability of a higher number of possible tracers in the inner regions has not improved so much the situation with the same data be reproducible in terms of different (and somewhat contrasting) scenarios because of the uncertainties on the stellar initial mass function (IMF).

In the classical Newtonian framework, dark matter (DM) dominates the outer galaxy regions (see, e.g., van den Bosch et al. 2003a, 2003b, 2007), but weighting its contribution in the inner regions where stellar matter plays a not negligible role is quite difficult. As pointed out by Mamon \& Lokas (2005a,b), the observational data claim for a dominant stellar component at a radius $ \lsim \, \ R_{eff}$, but such a conclusion heavily relies on the IMF choice, with the Salpeter (1995) and Chabrier (2001) as leading but not unique candidates. On large scales, the DM content has been found to be a strong function of both luminosity and mass \cite{Benson2000,MH02,vdeB07} with a different behaviour between faint and bright systems. Looking for a similar result for the DM content at $\sim  R_{eff}$ is quite controversial. On one hand, some authors \cite{Ger01,BSD2003} argue for no dependence. In contrast, other works \cite{Nap05,Cappellari06,cresc08a} do find that brighter galaxies have a larger DM content, while a flattening and a possible inversion of this trend for lower mass systems, similar to the one observed for late\,-\,type galaxies \cite{Persic93}, is still under analysis \cite{Nap05,cresc08a} with no conclusive result yet obtained. It is worth stressing that all these results rely on dynamical analysis, i.e. they are obtained by fitting a given model to the observed velocity dispersion data. As such, they are plagued by the mass\,-\,anisotropy degeneracy introducing an unknown systematic bias which is difficult to quantify. As an alternative, one can rely on gravitational lensing which directly probes the full mass content projected along the line of sight. In the strong lensing regime, the formation of Einstein rings allows to constrain the total mass projected within the Einstein ring independently of the model. If coupled to a measurement of the central velocity dispersion, this method gives interesting constraints on the mass profile in the inner regions. On one hand, dark matter mass fractions and their scaling with stellar mass and luminosity can be constrained assuming a model for the halo \cite{KT03,G09,Slacs1,Slacs2,T10,CT10} or the total mass profile \cite{CTMS09,G10}. On the other hand, one can also try to constrain the stellar IMF by making some assumptions on the dark matter content \cite{GC10}. In the opposite regime, galaxy\,-\,galaxy lensing allows to investigate the outer regions estimating the total mass\,-\,to\,-\,light ($M/L$) ratio \cite{GS02,H05,G07} although large samples are needed to get reliable results.

In spite of the quite large range spanned by their morphological and photometric properties, ETGs show several interesting correlations among their colors, luminosities, velocity dispersions, effective radii and surface brightness, the most famous example being the Fundamental Plane FP, a log\,-\,linear relation between the effective radius $R_{eff}$, the intensity $I_e = I(R_{eff})$, and the central velocity dispersion $\sigma_0$ (Djorgovski \& Davis 1987, Dressler et al. 1987, Bender et al. 1992, Burstein et al. 1997). This fundamental plane is usually parameterized as\,: $R_{eff} \propto \sigma_0^a I_e^b$ with $(a, b)$ predicted to be $(2, -1)$ if ETGs are homologous systems in virial equilibrium and with a constant $M/L$ ratio. The observed plane is, however, tilted with respect to the virial one since the different determinations of $(a, b)$, depending on the photometric band and the sample used, are always different from the virial values. Jorgensen et al (1996) first derived $a = 1.24 \pm 0.07$ and $b = -0.82 \pm 0.02$ from a set of 225 early\,-\,type galaxies in nearby clusters observed in the r\,-\,band. While this result is consistent with the original observations of Djorgovsky \& Davis (1987) and Dressler et al. (1987), it is nevertheless in striking contrast with the most recent determination relying on $\sim 9000$ ETGs observed within the framework of the Sloan Digital Sky Survey (SDSS). Using this large sample, Bernardi et al. (2003) have found $a = 1.49 \pm 0.05$ and $b = -0.75 \pm 0.01$, which are more similar to the K\,-\,band fundamental plane of Pahre et al. (1998). While the precise values of the FP coefficients are still debated, it is nevertheless clear that the observed FP is tilted with respect to the predicted plane. Such a tilt could be caused by a variation in the dynamical $M/L$ ratio for ETGs as a result of a varying dark matter fraction (e.g., Padmanabhan et al. 2004; Boylan\,-\,Kolchin et al. 2005, Cappellari et al. 2006, Tortora et al. 2009) or stellar population variations (e.g., Gerhard et al. 2001). Moreover, non\,-\,homology in the surface brightness profiles of elliptical galaxies (e.g., Graham \& Colless 1997; Trujillo et al 2004) may be an other explanation of the FP tilt\footnote{Hereafter, with an abuse of terminology, we will refer to the plane with coefficients $(a, b) = (2, -1)$ as the {\it virial FP} even if, as explained, deviations from this plane do not imply that ETGs are unvirialized systems.}.

All the above results have been obtained assuming that the classical Newtonian theory of gravity may be used also on galactic scales. However, the
outer regions of galaxies typically are in a low acceleration regime and in this regime Newtonian dynamics has never been experimentally tested. Motivated by this consideration, Milgrom (1983) proposed to modify Newton's second law of dynamics as $F = m g$, where the acceleration $g$ is now related to the Newtonian one $g_N$ as $g \mu(g/a_0) = g_N$. Here, $a_0$ is a new universal constant and $\mu(x)$ may be an arbitrary function with the properties $\mu(x >> 1) = 1$ and $\mu(x << 1) = x$. The theory thus obtained, referred to as {\it Modified Newtonian Dynamics} (MOND), provides flat rotation curves for spiral galaxies and, as a by\,-\,product, gives a theoretical interpretation of the empirically determined Tully\,-\,Fisher law. Since its beginnings, MOND has been widely tested as an alternative to dark matter with remarkable success on galaxy scales (see, e.g., Sanders \& McGaugh 2002, Milgrom 2008 and refs. therein) receiving a renewed interest after the proposal of a possible fully covariant relativistic formulation referred to as TeVeS \cite{TeVeS}.

Motivated by these successful results, we model ETGs as a single component system derived from the observed luminosity profile and check whether the resulting MOND dynamics is consistent with the data. To this end, assuming a constant anisotropy model for the velocity dispersion profile, we compute the aperture velocity dispersion $\sigma_{ap}$ and compare it with the observed value for a large sample of local ETGs. Fitting the $\sigma_{ap}$ data provides us a consistency check of MOND. Indeed, should the MOND predicted values be smaller than the observed one or ask for unrealistic anisotropy profile, one can argue that DM is still needed. Should, on the contrary, MOND overestimates $\sigma_{ap}$, one should reconsider the choice of the interpolating function $\mu(x)$ or the acceleration constant $a_0$. As a byproduct, this test also allows us to estimate the anisotropy profile which can then be used to infer how the theoretical FP looks like in the MOND framework. It is worth noting, however, that $\sigma_{ap}$ mainly probe the inner regions of the galaxy where dark matter is not expected to play a dominant role. As a consequence, although our analysis is definitely in the context of the MOND vs DM controversy, we do not expect to give a conclusive answer on this debate, but we nevertheless make a further step towards a possible solution.

The plan of the paper is as follows. In Section 2, we present the mass density profile adopted for describing the ETGs luminous component, while the derivation of the aperture velocity dispersion in the MOND framework is given in Section 3. We then show, in Section 4, how MOND predicts a FP\,-\,like relation for ETGs which we refer to as the {\it MOND Fundamental Plane} (MFP). Section 5 is then devoted to both motivating the MFP and testing the viability of MOND in reproducing the observed dynamics of ETGs, while the predicted MFP coefficients are given in Section 6 and compared to those of the observed FP. We finally summarize our results and discuss possible implications in Section 7, while in the Appendix A and B we briefly investigate the wavelength dependence of the MFP coefficients and discuss the impact of deviations from the main assumptions considered in the paper.

\section{The PS model}

In spite of their wide mass, size and luminosity range and the different chemical and stellar population characteristics, ETGs present a remarkable similarity in their photometric properties. As many studies show, \cite{CCD93,GC97,PS97}, their surface brightness is well described by the Sersic (1968) profile\,:

\begin{equation}
I(R) = I_e \exp{\left \{ - b_n \left [ \left ( \frac{R}{R_{eff}} \right )^{1/n} - 1 \right ] \right \}}
\label{eq: ir}
\end{equation}
with $R$ the cylindrical radius\footnote{Note that we have implicitly assumed that the intensity $I$ does not depend on the angular coordinates. Actually, the isophotes are not concentric circles, but rather ellipses with variable ellipticities and position angles so that $I = I(R, \varphi)$. However, in order to be consistent with our assumption of spherical symmetry of the three dimensional mass profile, we will neglect such an effect and, following a common practice, {\it circularize} the intensity profile considering circular isophothes with radii equal to the geometric mean of the major and minor axes.} on the plane of the sky and $I_e$ the luminosity intensity at the effective radius $R_{eff}$. The constant $b_n$ is determined by the condition that the luminosity within $R_{eff}$ is half the total luminosity. A very good analytical approximation is given by \cite{CB99}\,:

\begin{displaymath}
b_n = 2n - \frac{1}{3} + \frac{0.009876}{n} \ .
\end{displaymath}
The deprojection of the intensity profile in Eq.(\ref{eq: ir}) is straightforward under the hypothesis of spherical symmetry, but, unfortunately, the result turns out to be somewhat involved combinations of the unusual Meijer functions \cite{MC02}. In order not to deal with this cumbersome expression, we prefer to use the model proposed by Prugniel and Simien (1997, hereafter PS) whose three dimensional luminosity density reads\,:

\begin{equation}
j(r) = j_0 \left ( \frac{r}{R_{eff}} \right )^{-p_n} \exp{\left [ -b_n \left ( \frac{r}{R_{eff}} \right )^{1/n} \right ]}
\label{eq: jr}
\end{equation}
with

\begin{equation}
j_0 = \frac{I_0 b_n^{n (1 - p_n)}}{2 R_{eff}} \frac{\Gamma(2n)}{\Gamma[n (3 - p_n)]} \ .
\label{eq: jz}
\end{equation}
Here, $\Gamma(a)$ is the $\Gamma$ function, $I_0 = I(R = 0) = I_e {\rm e}^{b_n}$, while the constant $p_n$ is chosen so that the projection of Eq.(\ref{eq: jr}) matches a Sersic profile with the same values of $(n, R_{eff}, I_e)$. A useful fitting formula is given as \cite{Lima99,Metal01}\,:

\begin{displaymath}
p_n = 1.0 - \frac{0.6097}{n} + \frac{0.00563}{n^2} \ .
\end{displaymath}
In the following, we will be interested in the mass rather than luminosity density. Under the {\it light\,-\,traces\,-\,mass} assumption, the two quantities are immediately related as $\rho(r) = \Yst j(r)$ with $\Yst$ the stellar $M/L$ ratio. For later applications, it is convenient to define the following dimensionless quantity

\begin{equation}
\tilde{\rho}(\eta) = \frac{\rho(\eta)}{\rho_{eff}} = \eta^{-p_n} \exp{[-b_n (\eta^{1/n} - 1)]} \ ,
\label{eq: rhotilde}
\end{equation}
with $\eta = r/R_{eff}$ and

\begin{equation}
\rho_{eff} = \rho(\eta = 1) = \Yst j_0 {\rm e}^{-b_n} = \frac{M_{\star}}{4 \pi R_{eff}^3} \frac{b_n^{n(3 - p_n)}
{\rm e}^{-b_n}}{n \Gamma[n(3 - p_n)]} \ .
\label{eq: rhoeff}
\end{equation}
Here $M_{\star} = \Yst L_T$ is the total stellar mass with $L_T = 2 \pi n b_n^{-2n} {\rm e}^{b_n} \Gamma(2n) I_e R_{eff}^2$ the total luminosity of the projected Sersic profile. Because of the assumed spherical symmetry, it is only a matter of algebra to show that the cumulative mass profile
for the PS model reads\,:

\begin{equation}
M(r) = M_{\star} \ \frac{\gamma[n (3 - p_n), b_n \eta^{1/n}]}
{\Gamma[n (3 - p_n)]} \ ,
\label{eq: mr}
\end{equation}
where $\gamma(a,x)$ is the incomplete $\gamma$ function. It is useful to introduce the following scaled mass profile\,:

\begin{equation}
\tilde{M}(\eta) = \frac{M(\eta)}{M_{eff}} = \frac{\gamma[n(3 - p_n), b_n \eta^{1/n}]}{\gamma[n(3 - p_n), b_n]}
\label{eq: masstilde}
\end{equation}
with

\begin{equation}
M_{eff} = M(\eta = 1) =  M_{\star} \ \frac{\gamma[n (3 - p_n), b_n]} {\Gamma[n (3 - p_n)]} \ .
\label{eq: masseff}
\end{equation}
As a final remark, let us stress that both $M_{\star}$ and $M_{eff}$ may be determined from the measurement of the photometric parameters $(n, R_{eff}, I_e)$ provided that an estimate of the stellar \ML\ ratio \Yst\ is available (for instance, from the relation between \Yst\ and the
colours or from fitting the galaxy spectrum to stellar population synthesis models).

\section{Aperture velocity dispersion}

A widely used probe to constrain the model parameters is represented by the line of sight velocity dispersion luminosity weighted within a circular aperture of radius $R_{ap}$. This can be easily evaluated as\,:

\begin{equation}
\sigma_{ap}^2 = \frac{\int_{0}^{R_{ap}}{I(R) \sigma_{los}^2(R) R dR}} {\int_{0}^{R_{ap}}{I(R) R dR}}
\label{eq: defsigmaap}
\end{equation}
with $\sigma_{los}(R)$ the velocity dispersion projected along the line of sight. In order to compute this latter quantity, we first need to solve the Jeans equation for the radial velocity dispersion $\sigma_r$. Following Sanders (2000) and assuming spherical symmetry,
we write it as\,:

\begin{equation}
\frac{d[\rho(r) \sigma_r^2(r)]}{dr} + \frac{2 \beta(r)}{r} [\rho(r) \sigma_r^2(r)] = -\rho(r) g(r)
\label{eq: jeanseq}
\end{equation}
with $\beta(r) = 1 - \sigma_\theta^2/\sigma_r^2$ the anisotropy profile and $g(r)$ the acceleration law. In the classical Newtonian dynamics,
$g(r) = G M_{tot}(r)/r^2$ with $M_{tot}(r)$ the total (stellar + DM) mass. In the MOND framework, no DM is added so that $M_{tot}(r) = M(r)$ with $M(r)$ the stellar mass from Eq.(\ref{eq: mr}). Second (and most importantly), the acceleration $g(r)$ is obtained by solving

\begin{equation}
g(r) \mu \left [ \frac{g(r)}{a_0} \right ] = g_N(r)
\label{eq: mondgeq}
\end{equation}
with $\mu(x)$ the MOND interpolating function and $a_0 = 1.2 \times 10^{-10} \ {\rm m/s^2}$ the MOND acceleration scale. Without loss of generality, we conveniently define\,:

\begin{equation}
g(r) = \gamma_{MOND}(r) g_N(r)
\label{eq: defgammamond}
\end{equation}
where the functional expression of $\gamma_{MOND}(r)$ will depend on the mass model and the $\mu(x)$ function. Although $\mu(x) = x/\sqrt{1 + x^2}$ has been the first proposal to be tested with success \cite{SmG02}, recent analyses \cite{FB05,ZF06,FBZ07,SE07,Garry08b} seem to favour the {\it simple form} \cite{FB05}

\begin{equation}
\mu(x) = \frac{x}{1 + x} \ ,
\label{eq: simplemu}
\end{equation}
which we will adopt in the following. Using Eq.(\ref{eq: mr}) for the mass profile, inserting Eq.(\ref{eq: simplemu}) into Eq.(\ref{eq: mondgeq}) and solving for $g(r)$, we finally find\,:

\begin{equation}
2 \gamma_{MOND}(\eta) =
1 + \sqrt{1 + \frac{4 a_0}{a_{eff}} \frac{\eta^2 \gamma[n(3 - p_n), b_n]} {\gamma[n(3 - p_n), b_n \eta^{1/n}]}}
\label{eq: simplegamma}
\end{equation}
with

\begin{equation}
a_{eff} = \frac{G M_{eff}}{R_{eff}^2} = \frac{G M_{\star}}{R_{eff}^2} \ \frac{\gamma[n (3 - p_n), b_n]}
{\Gamma[n (3 - p_n)]} \ .
\label{eq: defaeff}
\end{equation}
The computation of $\sigma_{los}$ may now be performed along the same steps as for the Newtonian case (see, e.g, Mamon \& Lokas, 2005a,b) provided the following replacement rule

\begin{displaymath}
G M(r)/r^2 \rightarrow G M(r) \gamma_{MOND}(r)/r^2
\end{displaymath}
is applied. The final result turns out to be\,:

\begin{eqnarray}
I(R) \sigma_{los}^2(R) & = & 2 G M_{eff} \rho_{eff}/\Yst \\ ~ & \times &
\int_{\xi}^{\infty}{K\left ( \frac{\eta}{\xi}, \frac{\eta_a}{\xi} \right )
\frac{\tilde{\rho}(\eta) \tilde{M}(\eta) \gamma_{MOND}(\eta)}{\eta} d\eta} \nonumber
\label{eq: sigmalosgen}
\end{eqnarray}
with $\xi = R/R_{eff}$, $\eta_a = r_a/R_{eff}$ a scaled anisotropy radius and $K(\eta/\xi, \eta_a/\xi)$ a kernel function depending on the choice of
the anisotropy profile\footnote{For some $\beta(r)$ profiles, the kernel is analytic and can be retrieved from Appendix B of Mamon \& Lokas (2005b).}. There are not many constraints on what the correct anisostropy profile can be. Moreover, the results in the literature have all be obtained under the assumption of Newtonian gravity and dark matter so that extrapolating them to the MOND framework is not motivated. As a first approximation, we will therefore consider models with constant anisotropy, i.e. $\beta(r) = \beta$, where we remember that $\beta$ may range from $-\infty$ (for a system with fully tangential orbits) to $1$ (for radial orbits only). For the anisotropic PS model in a MOND framework, we finally get\,:

\begin{equation}
\sigma_{ap}^2 = \frac{4 \pi G \rho_{eff} M_{eff} R_{eff}^2} {\Yst L_2 \gamma[n(3 - p_n), b_n]}
\int_{0}^{\xi_{ap}}{{\cal{I}}_{los}(\xi, {\bf p}) \xi d\xi} \ ,
\label{eq: sigmaap}
\end{equation}
with

\begin{eqnarray}
{\cal{I}}_{los}(\xi, {\bf p}) & = & \int_{\xi}^{\infty}{\frac{\gamma_{MOND}(\eta)}{\eta^{1 + p_n} \exp{[b_n (\eta^{1/n} - 1)]}}
K\left ( \frac{\eta}{\xi}, \beta \right )} \nonumber \\
~ & \times & \ \ \frac{\gamma[n(3 - p_n), b_n \eta^{1/n}]}{\Gamma[n(3 - p_n)]} \ d\eta \ ,
\label{eq: defilos}
\end{eqnarray}

\begin{equation}
L_2 = 2 \pi n {\rm e}^{b_n} b_n^{-2n} I_e R_{eff}^2 \gamma(2n, b_n \xi_{ap}^{1/n}) \ ,
\label{eq: defl2}
\end{equation}
and ${\bf p}$ denotes the set of parameters on which the integral depends. In the FP studies, the aperture velocity dispersion is usually referred to
a circular aperture of radius $R_{ap} = R_{eff}/8$, i.e. $\xi_{ap} = 1/8$. Following common practice, we will set $\sigma_0 = \sigma_{ap}(\xi_{ap} = 1/8)$ and note that, because of Eqs.(\ref{eq: sigmaap})\,-\,(\ref{eq: defl2}), it is\,:

\begin{displaymath}
\sigma_0^2 = s(n, I_e, R_{eff}, \Yst, \beta)
\end{displaymath}
having noted that both $\rho_{eff}$ and $M_{eff}$ may be expressed as function of the photometric parameters $(n, R_{eff}, I_e)$ and the stellar $M/L$ ratio $\Yst$. Investigating the shape of the function $s({\bf p})$ will be the aim of the next section.

\section{The theoretical MOND FP}

It is just a matter of algebra to show that

\begin{equation}
\sigma_0^2 = \frac{G M_{\star}}{R_{eff}} \frac{b_n {\rm e}^{-b_n} \Gamma(2n)}
{n \Gamma[n(3 - p_n)]} \frac{\tilde{s}({\bf p_s})}{\gamma[2n, b_n (1/8)^{1/n}]}
\label{eq: sfp}
\end{equation}
with

\begin{equation}
\tilde{s}({\bf p_s}) = \int_{0}^{1/8}{{\cal{I}}_{los}(\xi, {\bf p_s}) \xi d\xi}
\label{eq: deftildes}
\end{equation}
and we have introduced the subset of parameters\,:

\begin{displaymath}
{\bf p_s} = (n, a_{eff}, \beta)
\end{displaymath}
with $a_{eff}$ is a function of  $(n, I_e, R_{eff}, \Upsilon_{\star})$ through $(M_{eff}, R_{eff})$. It is computationally convenient to use $a_{eff}/a_0$ as parameter when estimating $\tilde{s}({\bf p}_s)$ since this is the only way the full set of parameters enters the definition of the $\gamma_{MOND}(\eta)$ function in the integral.

We can now make use of the definition of average effective intensity to write\,:

\begin{displaymath}
M_{\star} = \Yst L_T = 2 \pi \Yst \langle I_e \rangle R_{eff}^2
\end{displaymath}
where $\langle I_e \rangle$ and $I_e$ are related by \cite{GD05}

\begin{equation}
\langle I_e \rangle = n b_n^{-2n} {\rm e}^{b_n} \Gamma(2n) I_e \ .
\label{eq: defiemedio}
\end{equation}
Note that, using $\langle I_e \rangle$, the parameter $a_{eff}$ now reads\,:

\begin{equation}
a_{eff} = \frac{2 \pi G \langle \Sigma_e \rangle \gamma[n (3 - p_n), b_n]} {\Gamma[n (3 - p_n)]}
\label{eq: defaeffbis}
\end{equation}
where we have defined the average mass surface density as\,:

\begin{equation}
\langle \Sigma_e \rangle = \Yst \langle I_e \rangle \ .
\label{eq: defsigmaeff}
\end{equation}
Because of Eq.(\ref{eq: defsigmaeff}), Eq.(\ref{eq: sfp}) can be seen as a relation among $(n, R_{eff}, \langle \Sigma_e \rangle)$ which can be easily solved by introducing logarithmic units as\,:

\begin{eqnarray}
\log{R_{eff}} & = & 2 \log{\sigma_0} - \log{\langle \Sigma_e \rangle} - \log{\tilde{s}(n, \langle \Sigma_e \rangle)} \nonumber \\
~ & - & \log{f(n)} - \log{(2 \pi G)}
\label{eq: reffsolve}
\end{eqnarray}
with

\begin{equation}
f(n) = \frac{b_n {\rm e}^{-b_n} \Gamma(2n)}{n \Gamma[n(3 - p_n)]} \ .
\label{eq: defeffen}
\end{equation}
Let us now assume that, over the parameter space actually covered by galaxies, the two terms $\tilde{s}(n, \langle \Sigma_e \rangle)$ and $f(n)$ may be well described as power law functions of their arguments. In such a case, Eq.(\ref{eq: reffsolve}) may be approximated as\,:

\begin{equation}
\log{R_{eff}} = a \log{\sigma_0} + b \log{\langle \Sigma_e \rangle} + c \log{n} + d
\label{eq: premfp}
\end{equation}
with $(a, b, c, d)$ parameters to be determined as described later. A guess for their values can be obtained by means of the following expected scaling relations\,:

\begin{displaymath}
\tilde{s}(n, \langle \Sigma_e \rangle) \propto n^{a_s} \langle \Sigma_e \rangle^{b_s} \ \ ,
\ \ f(n) \propto n^{a_n} \ \ ,
\end{displaymath}
so that one should find\,:

\begin{displaymath}
a = 2.0 \ \ , \ \ b = -(b_s + 1) \ \ , \ \ c = -a_n \ \ .
\end{displaymath}
Actually, deviations of $\tilde{s}(n, \langle \Sigma_e \rangle)$ and $f(n)$ from exact power laws may be, in principle, compensated introducing a dependence on $\sigma_0$ thus making $a$ deviating from 2. We have therefore left this parameter free in order to accomodate possible deviations from the canonical value induced by the change from the exact Eq.(\ref{eq: reffsolve}) to the approximated (\ref{eq: premfp}).

As a final step, we now have to remember that both the stellar $M/L$ ratio $\Yst$ \cite{PS96,Cappellari06,cresc08a} and the Sersic index $n$ \cite{CCD93,GG03,MamonLokas05a} may actually be correlated with the galaxy total luminosity $L$ (hereafter, we drop the label $T$ for convenience). In a first good approximation, we can write\,:

\begin{equation}
\log{\Yst} = \alpha_{\star} \log{L} + \beta_{\star} \ ,
\label{eq: mlvsl}
\end{equation}

\begin{equation}
\log{n} = \alpha_{\nu} \log{L} + \beta_{\nu} \ .
\label{eq: nvsl}
\end{equation}
Using then Eqs.(\ref{eq: defiemedio}) and (\ref{eq: defsigmaeff}) and putting together Eqs.(\ref{eq: premfp}), (\ref{eq: mlvsl}) and (\ref{eq:
nvsl}), we finally get\,:

\begin{equation}
\log{R_{eff}} = \alpha_{MFP} \log{\sigma_0} + \beta_{MFP} \log{\langle I_e \rangle} + \gamma_{MFP}
\label{eq: mfp}
\end{equation}
with

\begin{equation}
\left \{
\begin{array}{l}
\displaystyle{\alpha_{MFP} = \frac{a}{1 - 2(b \alpha_{\star} + c \alpha_{\nu})}} \\
~ \\
\displaystyle{\beta_{MFP} = \frac{(1 + \alpha_{\star}) b + c \alpha_{\nu}}{1 - 2(b \alpha_{\star} + c \alpha_{\nu})}} \\
~ \\
\displaystyle{\gamma_{MFP} = \frac{(b \alpha_{\star} + c \alpha_{\nu}) \log{(2 \pi)} + b \beta_{\star} + c \beta_{\nu} + d}
{1 - 2(b \alpha_{\star} + c \alpha_{\nu})}} \\
\end{array}
\right .
\label{eq: mfpcoeff}
\end{equation}
which defines what we call the {\it MOND Fundamental Plane}.

According to the above derivation, the MOND Fundamental Plane\footnote{It is worth noting that the problem of the FP in MOND has been first discussed in Sanders (2000). However, in that paper, Sanders did work out a galaxy model looking for a solution of the Jeans equations for a polytropic system, while, here, we start from what we observe. See, also, Sanders \& Land (2008) and Sanders (2010) for a similar approach.} (hereafter MFP) should be a
perfect plane with no scatter. Actually, there are three main sources of scatter. First, Eq.(\ref{eq: premfp}) is an approximation for the exact Eq.(\ref{eq: reffsolve}). Thus, we expect that this works more or less well depending on each galaxy. In other words, the difference $\Delta R_{eff}$ between the solutions of Eqs.(\ref{eq: reffsolve}) and (\ref{eq: premfp}) will be a function of $(n, R_{eff}, \langle \Sigma_e \rangle)$ and
hence will change from one galaxy to another introducing a scatter around the MFP. As a second issue, one should take into account that the $n$\,-\,$L$ and $\Yst$\,-\,$L$ correlations are affected by an intrinsic scatter that propagates to the MFP. Finally, although not explicitly denoted above, it is nevertheless clear from Eq.(\ref{eq: defilos}) that $\tilde{s}(n, \langle \Sigma_e \rangle)$ also depends on the anisotropy parameter $\beta$. Since this quantity changes from one galaxy to another, this will induce deviations from a fiducial MFP defined by setting $\beta$ to a reference value. The combination of all these effects make the galaxies scattering around the MFP thus giving it a non zero thickness.

\section{Testing the MOND scenario}

The previous sections have demonstrated how to derive a fundamental plane in the MOND framework. However, the above derivation heavily relies on two
main assumptions. First, we have assumed that the exact relation (\ref{eq: reffsolve}) may be replaced by the approximated expression (\ref{eq: premfp}). Second, we are implicitly assuming that one is able to reproduce the dynamics of ETGs in the MOND framework without introducing any further dark matter component. In order to test this latter hypothesis against observational data, one can try to fit the observed aperture velocity dispersion in a large sample of local ETGs with well measured photometric properties. In such a case we can assume the three photometric quantities $(n, R_{eff}, \langle I_e \rangle)$ to be known so that imposing that the observed and theoretically predicted $\sigma_{ap}^2$ are equal gives a constraint in the parameter space $(\Yst, \beta)$. We can moreover resort to stellar population synthesis codes to infer the stellar $M/L$ ratio $\Yst$ from the galaxy colors thus finally ending up with an estimate of the anisotropy parameter $\beta$. Note that such a sample is also useful to test the first assumption and suggests what is the region of the parameter space $(n, R_{eff}, \langle \Sigma_e \rangle)$ we have to explore to see whether our approximation holds for realistic ETGs systems.

\subsection{The data}

As a first step to carry on the approach detailed above, one has to assemble a sample of ETGs as large as possible. To this aim, we have started from the NYU Value\,-\,Added Galaxy Catalog (hereafter, VAGC) which is a cross\,-\,matched collection of galaxy catalogs maintained for the study of galaxy formation and evolution \cite{VAGC} and mainly based on the SDSS data release 6 \cite{DR6}. Among the vast amount of available data, we use the {\it low\,-\,redshift} (hereafter, lowZ) catalog of galaxies with estimated comoving distances in the range $10 < D < 150 \ h^{-1} {\rm Mpc}$. We refer the reader to Blanton et al. (2005) and the VACG website\footnote{{\tt http://cosmo.nyu.edu/blanton/vagc/}} for details on the compilation of the catalog. Note that the lowZ catalog is actually updated only to the fourth \cite{DR4} rather than the sixth SDSS data release thus covering an effective survey area of 6670 square degrees.

From the lowZ catalog, we remove all the galaxies with unmeasured $\sigma_0$. This first cut leaves us with 43312 out of 49968 objects with magnitudes in the five SDSS filters $u' g' r' i' z'$. Based on the available data, we assemble an ETGs sample by imposing the following selection criteria\,:

\begin{enumerate}

\item{$2.5 < n < 5.5$ with $n$ the Sersic index in the $i'$ band and the upper end dictated by the code limit $n = 6.0$;}

\item{$R_{90}/R_{50} > 2.6$ \cite{Shima01} with $R_{f}$ the Petrosian radii containing $f\%$ of the total luminosity;}

\item{$\sigma_0 > 70 \ {\rm km/s}$ since the measurement of the velocity dispersion for such small mass system could be unreliable as explained in Bernardi et al. (2005);}

\item{$(g' - r')_{-} \le g' - r' \le (g' - r')_{+}$ with $(g' - r')_{\pm} = p M_r + q {\pm} \delta$, $M_r$ the absolute magnitude in the $r'$ filter and the parameters $(p, q, \delta)$ have been tailored from Fig.\,2 in Bernardi et al. (2005) where a different ETG sample has been extracted from the SDSS DR2 \cite{DR2}.}

\end{enumerate}
The final sample thus obtained contains 9046 galaxies out of an initial catalog containing 49968 objects. It is worth noting that most of the rejected objects have been excluded by the first three cuts (retaining only 9105 entries), while the fourth cut only removes 59 further galaxies. This is reassuring since the last cut is somewhat qualitative and based on a different set of selection criteria \cite{B05}. We then use the data reported in the lowZ catalog for the galaxies in the above sample to collect the quantities of interest. In particular, the average effective intensity $\langle I_e \rangle$ is given by\,:

\begin{equation}
\langle I_e \rangle = 10^{-6} {\times} \frac{{\rm dex}[({\cal{M}}_t - {\cal{M}}_{\odot})/2.5]}
{2 \pi R_e^2}
\label{eq: ieest}
\end{equation}
with ${\rm dex}(x) \equiv 10^{x}$, ${\cal{M}}_t$ the galaxy absolute magnitude corrected for extinction, evolution and cosmological dimming, ${\cal{M}}_{\odot}$ the Sun absolute magnitude in the given filter\footnote{We use ${\cal{M}}_{\odot} = (5.82, 5.44, 4.52, 4.11, 3.89)$ for the $u' g' r' i' z'$ filters respectively as evaluated from a detailed Sun model reported in {\tt www.ucolick.org/$\sim$cnaw/sun.html}}, while $R_{eff}$ is here expressed in $kpc$ rather than $arcsec$. To this aim, we simply use\,:

\begin{displaymath}
R_{eff}({\rm kpc}) = R_{eff}(arcsec) {\times} D_A(z)/206265
\end{displaymath}
with $D_A(z)$ the angular diameter distance in Mpc.

The velocity dispersion reported in the lowZ catalog is measured within a fixed aperture $R_{SDSS} = 1.5 \ arcsec$, while $\sigma_0$ entering the MFP refers to an aperture of radius $R_{ap} = R_{eff}/8$. To correct for this offset, we follow J$\o$rgensen et al. (1995, 1996) and set\,:

\begin{equation}
\sigma_0^{obs} = \sigma_0^{lowZ} {\times} \left ( \frac{R_{SDSS}}{R_{eff}/8} \right )^{0.04}
\label{eq: sigmacorr}
\end{equation}
with $\sigma_0^{lowZ}$ the value in the catalog and $R_{eff}$ in $arcsec$ here.

\subsection{Estimate of the stellar $M/L$ ratio}

The lowZ catalog galaxies have been observed in five photometric bands so that we can use the color information in order to infer their stellar $M/L$ ratios \cite{cresc08a}. To this aim, we start assembling a library of synthetic stellar population models obtained through the {\tt Galaxev} code \cite{BC03} varying the age of the population, its metallicity and time lag of the exponential star formation rate and assuming a Chabrier (2001)
initial mass function (IMF). Then, we use the tabulated $(u', g', r', i', z')$ apparent magnitudes (corrected for extinction) of each ETG to fit the above library of spectra (suitably redshifted to ETG redshift) to the colours, thus getting the estimates of $\Yst$ for each galaxy in the catalog. Note that these values may be easily scaled to a Salpeter (1955) or Kroupa (2001) IMF by multiplying by $1.8$ or $1.125$ respectively so that we can
explore other IMF choices\footnote{While this is correct for a Salpeter IMF, since it differs from a Chabrier IMF only for the low mass slope and predict very similar colours, for a Kroupa IMF this is not strictly true, but we could assume the scale factor above as a good approximation.}. We finally get an estimate of the total stellar mass $M_{\star}$ simply as $\Yst L_T$ with the total luminosity $L_T$ taken from the catalog itself after correcting for the loss of flux due to the SDSS use of Petrosian magnitudes. Because of the errors on the colours, our \Yst\ estimates are affected by uncertainties difficult to evaluate. As a possible way out, one can use a Monte Carlo\,-\,like procedure generating a set of colours from a Gaussian distribution centred on each mean colour and standard deviation equal to the colour uncertainty. Fitting the colours thus obtained to the synthetic spectra for each realization, one could generate a distribution of fitted parameters and take the median and median scatter of such a distribution as an estimate of \Yst\ and its uncertainty (see Tortora et al. 2009 for further details). We have tested that such a procedure gives errors on \Yst\ of order $10\%$, but applying this method to the full sample is too time demanding so that we prefer to neglect this source of uncertainty. Should we have used our data to constrain the parameters of each single galaxy, this choice could have lead to an underestimation of the errors. However, we are here mainly interested in the statistical properties of the ensemble rather than fitting each individual galaxy. The uncertainty on each galaxy model parameters simply shifts the position of the galaxy in the parameters space, but does not alter the full distribution. Therefore, we are confident that neglecting the error on \Yst\ has no impact on our results.

\subsection{Motivating the MFP}

Having estimated the stellar $M/L$ ratio $\Yst$, we are now able to derive for each galaxy in the catalogue the parameters $(n, R_{eff}, \langle \Sigma_e \rangle)$ we need to compute the central velocity dispersion $\sigma_0$. We first evaluate the function $\tilde{s}({\bf p}_s)$ over a finely spaced grid in $(\log{n}, \log{(a_{eff}/a_0)}, \beta)$. Considering typical values for the galaxies in the sample, we set the grid limits as\,: $2.0 \le n \le 6.0$ and $-2.5 \le \log{(a_{eff}/a_0)} \le 2.5$. Since we have no a priori information on what values the anisotropy parameter can take in a MOND scenario, we conservatively allow $\beta$ to run in the range $(0, 1)$ thus considering the full range for radial anisotropy.

Having thus computed $\tilde{s}({\bf p_s})$, we can then resort to Eq.(\ref{eq: sfp}) to estimate $\sigma_0$ over a finely spaced grid in $(n, \log{R_{eff}}, \log{\langle \Sigma_e \rangle})$ for different values of $\beta$. For each model in the grid, we then solve Eq.(\ref{eq: reffsolve}) and finally fit these values using Eq.(\ref{eq: premfp}) to estimate the coefficients $(a, b, c, d)$ as function of the anisotropy parameter. In order to quantify the quality of the approximation, we also estimate $\Delta_{rms}$ with $\Delta = \log{R_{eff}(ex)} - \log{R_{eff}(fit)}$ and $R_{eff}(ex)$, $R_{eff}(fit)$ the exact and approximated solutions.

It turns out that Eq.(\ref{eq: premfp}) indeed works quite well with $0.06 \le \Delta_{rms} \le 0.09$ and $\Delta$ values that are actually quite smaller than the rms one over most of the parameter space explored\footnote{As an alternative way of quantifying the goodness of the approximation, one can note that the quantity $\tilde{\Delta} = 1 - \log{R_{eff}(fit)}/\log{R_{eff}(ex)}$ is of order $0.01\%$ over most of the parameter space explored.}. Moreover, there is no trend with any of the parameters so that we can conclude that replacing the exact solution Eq.(\ref{eq: reffsolve}) with the approximated one Eq.(\ref{eq: premfp}) does not introduce any bias thus giving a theoretical support to our derivation of the MFP.

Such a test makes also possible to infer the values of the $(a, b, c, d)$ coefficients and how they depend on the anisotropy parameter. We first note that the use of the approximated formula has made the coefficients $(a, b)$ to deviate from the values $(2, -1)$ one should have inferred from Eq.(\ref{eq: reffsolve}) in the Newtonian case. While this is only a marginal difference for $a$, the effect is quite important for $b$. Moreover, it is clear that, although trends of all the coefficients with the anisotropy parameter are clearly detected, the variation may be essentially neglected for $a$. On the other hand, both $b$ and $c$ significantly depend on $\beta$ so that we expect a scatter in the MFP coefficients $(\alpha_{MFP}, \beta_{MFP})$ due to the variation of the anisotropy parameter from one galaxy to another. We finally stress that the dependence of $d$ on $\beta$ does not have any impact on the slope of the MFP, but introduces a scatter in the zeropoint $\gamma_{MFP}$ which translates into a scatter of galaxies around the best fit plane.

One could wonder whether these results depend on the adopted MOND interpolating function. To this aim, we have repeated the above analysis considering the standard form for $\mu(x)$ hence setting

\begin{equation}
2 \gamma_{MOND}^2(\eta) =
1 + \sqrt{1 + \left ( \frac{2 a_0}{a_{eff}} \frac{\eta^2 \gamma[n(3 - p_n), b_n]}
{\gamma[n(3 - p_n), b_n \eta^{1/n}]} \right )^2} \ .
\label{eq: standgamma}
\end{equation}
in Eq.(\ref{eq: defilos}). We find that the values of $(a, b, c)$ are essentially the same thus suggesting a very weak dependence of the theoretical MFP coefficients on the particular $\mu(x)$ function considered. This can be qualitatively explained noting that $\sigma_{0}$ is evaluated inside a very small aperture. Although formally the integral entering ${\cal{I}}_{los}(\xi)$ is defined along the full line of sight, integrating it over the very inner region of the galaxy, where $\gamma_{MOND}(\eta) \sim 1$, makes the details of this function unimportant and explains why the choice of $\mu(x)$ has such a small impact on MFP coefficients.

\subsection{Matching the data}

While the above discussion shows that the derivation of the MFP is theoretically well founded, we still have to verify that MOND is indeed able to match the dynamics of ETGs without resorting to dark matter. For each galaxy in the catalogue, we have both the photometric quantities $(n, R_{eff}, \langle I_e \rangle)$ and the stellar $M/L$ ratio $\Yst$ so that, using Eq.(\ref{eq: sfp}), we can compute the aperture velocity dispersion as a function of the anisotropy parameter only. Matching the observed $\sigma_{ap}$ to the theoretically predicted one makes then possible to constrain the constant anisotropy parameter $\beta$. Note that, to this end, we directly use the observed $\sigma_{ap}$, while $\sigma_0$ is used when considering
the MFP. Two considerations motivated this choice. First, we have estimated $\sigma_0$ converting $\sigma_{ap}$ through Eq.(\ref{eq: sigmacorr}). Actually, this relation has been derived assuming the validity of Newtonian gravity\footnote{Indeed, Eq.(\ref{eq: sigmacorr}) can be derived by assuming a galaxy model and the anisotropy profile and then computing the ratio $\sigma_{obs}/\sigma_0^{lowZ}$ as function of the mass and anisotropy parameters. This is indeed how J$\o$rgensen et al. (1995, 1996) derived their relation under the assumption that Newtonian gravity is correct.} , while our analysis is in the MOND framework. Although it is likely that Eq.(\ref{eq: sigmacorr}) still applies in the MOND case being $\sigma_0$ evaluated within the $a/a_0 >> 1$ region, it is a more conservative choice not to use any correction and then checking a posteriori the validity of Eq.(\ref{eq: sigmacorr}). As a second motivation, we note that $\sigma_0$ is evaluated within $R_{eff}/8$ where baryons are likely the dominant contribution and the Newtonian regime for acceleration holds. On the contrary, $R_{ap}/R_{eff}$ (with $R_{ap} = 1.5 \ arcsec$) spans the $95\%$ confidence range $(0.16, 1.58)$ with $R_{ap}/R_{eff} = 0.54$ as median value. Kinematical analysis in the Newtonian framework suggest that the dark matter mass fraction within $\sim 1 R_{eff}$ is of order $30\%$ (for a Salpeter IMF). We therefore expect that $\sigma_{ap}$ is more suited than $\sigma_0$ to test the validity of MOND since it refers to a region where DM may play a not negligible role.

A further remark is in order here before the fitting analysis. First, both the photometric quantities and the stellar $M/L$ ratio are known with an error and the same holds for the observed $\sigma_{ap}$. In order to take this into account, one could randomly generate a large ensemble of values for $(n, R_{eff}, \langle I_e \rangle, \Yst, \sigma_{ap})$ according to a multinormal distribution with central vector and covariance matrix set by the observed values. Solving for $\beta$ each time, we can finally estimate $\beta$ and its error from the distribution thus obtained. Because of the large ETGs sample we are dealing with, such a work is computationally quite expensive. However, we are here mainly interested in checking whether MOND is able to fit the data rather than deriving the exact value of the anisotropy parameter for every single galaxy. Moreover, we need the value of $\beta$ or $\log{\eta_a}$ to estimate $(\alpha_{MFP}, \beta_{MFP})$ and the scatter in the MFP so that we are interested in its probability distribution function. For these reasons, we will neglect the errors on the observed quantities and simply take their central values for each galaxy in the sample.

It is worth stressing that the shape of probability distribution for $\beta$ depends on the adopted filter, because both $\Yst$ and the photometric quantities $(n, R_{eff}, \langle I_e \rangle)$ are wavelength dependent. Moreover, the sample of galaxies we will fit is not the same for all the filters. In fact, for a given galaxy, the Sersic index $n$ changes from one filter to another so that it is possible that a galaxy with $n(i') > 2.5$ has $n(f) < 2.0$ in an another filter $f$, thus looking more similar to a disky system rather than an ETG one. Moreover, because of code failures or
observational problems, some galaxies in the $u'$ and $z'$ filters may have $n > 5.5$ or an unmeasured absolute magnitude so that they have to be rejected from the sample. We stress, however, that more than $90\%$ of the galaxies in the starting catalog are present in all the subsamples used to determine the $\beta$ distribution in the different filters. Here, we discuss only the results for the fit in the $i'$ filter thus using all the galaxies in the sample. The analysis in the remaining bands is similar.

\subsubsection{Results for constant anisotropy models}

Since both the photometric parameters and the stellar $M/L$ ratio have been fixed, we can straightforwardly solve the equation $\sigma_{ap,obs} = \sigma_{ap}(n, R_{eff}, \langle \Sigma_e \rangle, \beta)$ with respect to $\beta$ for each individual galaxy and then look at the distribution of the results. Such a procedure successfully works for all the galaxies in the sample with $\beta$ values definitely pointing towards a strong radial anisotropy. This result is not surprising because, as well known in Newtonian gravity, radial anisotropy helps to lower the need for dark matter in the inner region of elliptical galaxies \cite{Ger01,deL08}. For instance, De Lorenzi et al. (2008), using the NMAGIC code to fit the observed velocity dispersion profile of NGC 4697, have found evidence of a mildly varying anisotropy profile with $\beta \simeq 0.3$ in the inner regions. In the MOND framework, only part of the DM contribution is provided by the effect of the $\gamma_{MOND}$ function entering $\sigma_{ap}$, while radial anisotropy supplies the remaining term needed to match the observed velocity dispersion.

While radially anisotropic models are able to match the data, their physical consistency is not a priori guaranteed. In order to be self consistent, our density profile should be derived from a distribution function of the form $L^{-2 \beta} f_E(E)$ with $L$ the specific angular momentum and $f_E(E)$ a function of the binding energy $E$. The function $f_E(E)$ has to be positive definite in order the model to be physically meaningful. Necessary conditions for the consistency of anisotropic models in the Newtonian framework have been derived in the literature \cite{CP92,AE06,CMdZ09,CM10}. Some of them may be extended also to MOND models taking into account that, for this purpose, MOND models behave like Newtonian ones in an external effective potential $\Psi_{eff}$ \cite{SN10}. This latter can be derived by demanding that the MOND acceleration $g(r)$ entering Eq.(\ref{eq: jeanseq}) equals $d\Psi_{eff}/dr$, i.e. one must integrate (numerically) the equation $d\Psi_{eff}/dr = \gamma_{MOND}(r) G M(< r)/r^2$ with the condition that $\Psi_{eff}(r)$ vanishes at infinity.

In the Newtonian framework, An \& Evans (2006) have shown that, if the potential and the density are self consistently related, a necessary condition for the positivity of the DF of constant anisotropy models is $\gamma \ge 2 \beta$, where $\gamma = - d\ln{\rho}/d\ln{r}|_{r = 0}$ is the logarithmic slope of the density profile at the centre. Ciotti \& Morganti (2010) have however shown that this result can be extended to models with an external potential which is a case MOND can be compared to. We can therefore use the An \& Evans result for the consistency of MOND constant anisotropy models.  For the PS case, it is $\gamma = p(n)$ so that $\beta < p(n)/2$ must be taken as an upper limit on $\beta$ to have physically consistent models.

In order to take the above limit into account, we adjust both $\beta$ and an effective $M/L$ ratio $\kappa \Yst$, where $\kappa$ accounts for deviations from the stellar $M/L$ computed assuming a universal Chabrier IMF. To this end, for each galaxy in the sample, we determine $\kappa$ varying $\beta$ over the range $(-0.5, \beta_{max})$ with $\beta_{max} = p(n)/2$ and finally take as our best estimate the one with the smallest value of $|1 - \kappa|$. Note that, in this way, we get models that are both physically consistent and with only minor deviations from the initial recipe for the stellar $M/L$ ratio. Imposing $0.75 \le \kappa \le 2.25$ for the selected solution (see later for the motivation of such a selection criterium), we find that $92\%$ of the galaxies may be successfully fitted with this procedure. Specifically, from the $\beta$ and $\kappa$ distributions, we find\,:

\begin{displaymath}
\langle \beta \rangle = 0.07 \ \ , \ \
\beta_{med} = 0.18 \ \ ,
\end{displaymath}
\begin{displaymath}
{\rm 68\% \ CL \ :} \ \ (-0.40, 0.28) \ \ , \ \
{\rm 95\% \ CL \ :} \ \ (-0.45, 0.37) \ \ ,
\end{displaymath}
\begin{displaymath}
\langle \kappa \rangle = 0.98 \ \ , \ \
\kappa_{med} = 1.00 \ \ ,
\end{displaymath}
\begin{displaymath}
{\rm 68\% \ CL \ :} \ \ (0.96, 1.02) \ \ , \ \
{\rm 95\% \ CL \ :} \ \ (0.80, 1.03) \ \ .
\end{displaymath}
It is worth wondering whether the spread in $\kappa$ can be fully ascribed to variations in the stellar population properties. Indeed, the estimate of $\Yst$ from the galaxy colours depend on the details of the stellar population synthesis code we have used. There are many ingredients entering this code so that it is not fully unrealistic to expect that they can change from one galaxy to another, while here we have assumed them to be universal. For instance, a simple recipe to increase $\Yst$ is to change the IMF from the Chabrier one, that we have used here, to the Salpeter one. It is worth noting that there is indeed still an open debate on what the IMF actually is. On one hand, direct star counts and observations (mainly, in the Milky Way) point towards a Chabrier (or Kroupa) IMF, but our Galaxy is a spiral. On the other hand, gravitational lensing (see, e.g., Treu et al. 2010) and studies of the central DM fraction in local ETGs (Napolitano et al. 2010) suggest that also a Salpeter IMF can reconcile data and observations for ETGs provided a NFW model is assumed for the dark matter halo. Motivated by these contrasting results, one can not exclude an IMF varying with luminosity \cite{RC93,cresc08a} or a correlation between the timescale of exponential star formation rate and the luminosity. Moreover, the metallicity and the dust content may also change on a case\,-\,by\,-\,case basis. Considering a conservative $\sim 25\%$ uncertainty on $\Yst$ for a given IMF and that $\Yst$ has to be scaled by a factor 1.8 when replacing the Chabrier IMF with the Salpeter one, we finally get $0.75 \le \kappa \le 2.25$ as a range for this parameter. Should $\kappa$ be outside this (conservative) range, one should conclude that dark matter is needed to fill the gap between stellar and dynamical mass. Needless to say, the results we get for $\kappa$ are fully consistent with the statistical uncertainty on the estimated $\Yst$. While this outcome is expected by construction, the fact that we are able to find a physically acceptable value of $\beta$ makes us argue in favour of MOND being able to fit the velocity dispersion data with no dark matter provided a reasonable amount of anisotropy is allowed.

It is worth noting that the confidence ranges for $\beta$ are strongly asymmetric. Actually, $75\%$ of the fitted models have $\beta \ge 0$ with the remaining $25\%$ giving rise to a long flat tail towards negative $\beta$. If we simply cut our sample only retaining isotropic and radially anisotropic solutions, we find that the velocity dispersion may be fitted for $70\%$ of the galaxies. Actually, this fraction is still larger if we repeat the above analysis, but now imposing $\beta \ge 0$. The fit turns out to be successful for $87\%$ of the sample giving the following values characterizing the $\beta$ and $\kappa$ distributions\,:

\begin{displaymath}
\langle \beta \rangle = 0.17 \ \ , \ \
\beta_{med} = 0.20 \ \ ,
\end{displaymath}
\begin{displaymath}
{\rm 68\% \ CL \ :} \ \ (0.10, 0.30) \ \ , \ \
{\rm 95\% \ CL \ :} \ \ (0.0, 0.37) \ \ ,
\end{displaymath}
\begin{displaymath}
\langle \kappa \rangle = 0.98 \ \ , \ \
\kappa_{med} = 1.00 \ \ ,
\end{displaymath}
\begin{displaymath}
{\rm 68\% \ CL \ :} \ \ (0.95, 1.01) \ \ , \ \
{\rm 95\% \ CL \ :} \ \ (0.78, 1.03) \ \ .
\end{displaymath}
The spread in $\beta$ is now reduced with no statistical meaningful shift of the median value, while the shift in the mean is a consequence of the more symmetric distribution. The large fraction of successfully fitted galaxies then leads us to conclude that a mild radial anisotropy allows us to get physically consistent models able to fit the data with no need for dark matter in a MOND framework.

It is interesting to note that $\beta$ does not correlate neither with the photometric parameters $(n, R_{eff}, \langle I_e \rangle)$ nor with the total luminosity $L$. Such a result has an important outcome. Since the MFP coefficients $(\alpha_{MFP}, \beta_{MFP}, \gamma_{MFP})$ given by Eq.(\ref{eq: mfpcoeff}) depend on $\beta$ through $(a, b, c, d)$, a strong correlation of $\beta$ with $L$ could generate a correlation between $(\alpha_{MFP}, \beta_{MFP}, \gamma_{MFP})$ and $L_T$ and one should find different MFPs for different luminosity bins. However, this effect is likely to be weak because $(a, b, c, d)$ are actually weak function of $\beta$; moreover, we need to have an MFP independent of luminosity because the observed FP is approximately unique for all the ETGs.

As discussed in Appendix B, a correlation of $\kappa$ with $L$ could change the MFP coefficients. On the one hand, such a correlation is expected noting that the dynamical $M/L$ ratio $\Upsilon_{dyn} = \kappa \Yst$ is found to correlate with luminosity when fitting data in the Newtonian\,+\,dark matter framework. Surprisingly, we do not find any scaling relations between $\kappa$ and $L$. Indeed, this is a consequence of how we select the best fit model for each galaxy. Let us suppose that a relation like $\kappa = \kappa_s (L/L_s)^{\delta}$ exist for models with $\beta$ fixed. Depending on $(\kappa_s, L_s, \delta)$ and the galaxy luminosity, one could then obtain a value of $\kappa > 2.25$. But we have chosen as best fit model the one having as small a value as possible of $|1 - \kappa|$ so that, should $\kappa$ become too large, our algorithm would change $\beta$ (and hence $\kappa$) to minimize $|1 - \kappa|$. As a consequence, unless the correlation is quite strong, our selection criterim washes out a possible scaling of $\kappa$ with $L$. We stress, however, that this is not a limitation of our analysis, but rather a consequence of consistently working in a MOND framework.

\subsubsection{MOND vs Newtonian gravity}

The above results convincingly show that MOND is able to fit the aperture velocity dispersion in ETGs without the need for additional DM. Actually, the data we are using mainly probe the region $R < R_{eff}$ which is not in the deep MOND regime $(a << a_0)$. Indeed, for the galaxies in the sample, the median value of $\log{(a_{eff}/a_0)}$ is 0.56, while the $68\%$ confidence range is given by $0.21 \le \log{(a_{eff}/a_0)} \le 1.87$, i.e. $1.6 \le a_{eff}/a_0 \le 7.4$. We are therefore in an intermediate regime where the total acceleration $a$ is roughly of the same order of magnitude as $a_0$. We therefore expect that it is possible to fit the same data without DM also in the Newtonian regime. To this end, we have used the same equations as above setting $\gamma_{MOND}(r) = 1$ to recover the Newtonian formulae. Moreover, we have still to take into account the upper limit on $\beta$ so that we use the same procedure as above adjusting the anisotropy parameter and the ratio $\kappa$ between the dynamical and the stellar $M/L$ ratio.

Imposing the same cut on $\kappa$, we find that $\sim 80\%$ of the galaxies can be successfully fitted, a fraction smaller than in the MOND case, but still satisfactorily large. It is, however, interesting to note that, for most of the galaxies, the value of $\beta$ is quite close to the upper limit, while typical $\kappa$ values are larger than in the MOND case. Indeed, we get\,:

\begin{displaymath}
\langle \beta \rangle = 0.33 \ \ , \ \
\beta_{med} = 0.40 \ \ ,
\end{displaymath}
\begin{displaymath}
{\rm 68\% \ CL \ :} \ \ (0.02, 0.42) \ \ , \ \
{\rm 95\% \ CL \ :} \ \ (0.0, 0.44) \ \ ,
\end{displaymath}
\begin{displaymath}
\langle \kappa \rangle = 1.35 \ \ , \ \
\kappa_{med} = 1.29 \ \ ,
\end{displaymath}
\begin{displaymath}
{\rm 68\% \ CL \ :} \ \ (1.00, 1.79) \ \ , \ \
{\rm 95\% \ CL \ :} \ \ (0.81, 2.13) \ \ .
\end{displaymath}
As it is apparent from the $\kappa$ confidence ranges, one must increase the $M/L$ ratio up to very large values that, while still marginally consistent with deviations from the stellar one because of a different IMF, may be also easily interpreted in terms of a non negligible dark matter content. Needless to say, this is consistent with expectations since we know that, unless an unreasonably high radial anisotropy is introduced, Newtonian gravity can not explain ETGs dynamics without the boost provided by a dark matter halo.

It is worth comparing the values of $(\beta, \kappa)$ obtained in the MOND and Newtonian cases for each galaxy. We find\,:

\begin{displaymath}
\langle (1 + \beta_M)/(1 + \beta_N) \rangle = 0.85 \ \ , \ \
\beta_{med} = 0.86 \ \ ,
\end{displaymath}
\begin{displaymath}
{\rm 68\% \ CL \ :} \ \ (0.76, 0.95) \ \ , \ \
{\rm 95\% \ CL \ :} \ \ (0.71, 1.00) \ \ ,
\end{displaymath}
\begin{displaymath}
\langle \kappa_M/\kappa_N \rangle = 0.64 \ \ , \ \
\kappa_{med} = 0.67 \ \ ,
\end{displaymath}
\begin{displaymath}
{\rm 68\% \ CL \ :} \ \ (0.41, 0.85) \ \ , \ \
{\rm 95\% \ CL \ :} \ \ (0.24, 0.93) \ \ .
\end{displaymath}
These values clearly show that Newtonian models have a larger radial anisotropy and $\kappa$ value than MOND ones in agreement with the expecations. Indeed, for given model parameters, the boost to the Newtonian velocity dispersion provided by the $\gamma_{MOND}(r)$ factor reduces the need for both radial anisotropy and deviations from the fiducial stellar $M/L$. Not surprisingly, we find that $\kappa_M/\kappa_N$ is correlated with $\log{(a_{eff}/a_0)}$ so that the smaller is this quantity, the smaller is $\kappa_M/\kappa_N$. This is exactly what one can anticipate noting that the smaller is $\log{(a_{eff}/a_0)}$, the larger is the region of the galaxy where the MOND regime applies so that the smaller is the value of $\kappa_M$ needed to fit the data for the same value of the anisotropy parameter.

\section{The MFP coefficients}

Eq.(\ref{eq: mfpcoeff}) makes it possible to estimate the MFP parameters\footnote{Hereafter, we will parameterize the MFP through the two slope related quantities $(\alpha_{MFP}, \beta_{MFP})$ without considering anymore the zeropoint $\gamma_{MFP}$.} $(\alpha_{MFP}, \beta_{MFP})$ provided the values of $(\alpha_{\star}, \alpha_{\nu})$ are given and the anisotropy profile parameter has been set so that $(a, b, c)$ may be computed. The analysis in the previous section has demonstrated that MOND is able to fit the measured aperture velocity dispersion of the assembled ETGs sample and has enabled us to determine the distribution of $\beta$ for the galaxies in our sample. In order to compute the MFP coefficients, we first need to determine $(\alpha_{\star}, \alpha_{\nu})$ as we discuss in the following subsection.

\begin{table*}
\caption{Calibration parameters and intrinsic scatter for the luminosity scaling correlations
in Eqs.(\ref{eq: mlvsl}) and (\ref{eq: nvsl}) between $\log{L}$ and $\log{\Yst}$ and $\log{n}$
respectively. Columns are as follows\,: 1. filter id; 2. maximum likelihood parameters $(\alpha_{\star},
\beta_{\star}, \sigma_{int})$ for the $\log{L}$\,-\,$\log{\Yst}$ relation; 3, 4. median value and $68$
and $95\%$ confidence ranges for $(\alpha_{\star}, \sigma_{int})$; 5. maximum likelihood parameters $(\alpha_{\nu},
\beta_{\nu}, \sigma_{int})$ for the $\log{L}$\,-\,$\log{n}$ relation; 6., 7. median value and $68$ and
$95\%$ confidence ranges for $(\alpha_{\nu}, \sigma_{int})$.}
\begin{center}
\begin{tabular}{|c|c|c|c|c|c|c|}
\hline
Id & $(\alpha_{\star}, \beta_{\star}, \sigma_{int})_{ml}$ &
$(\alpha_{\star})_{-1\sigma \ -2\sigma}^{+1\sigma \ +2\sigma}$ &
$(\sigma_{int})_{-1\sigma \ -2\sigma}^{+1\sigma \ +2\sigma}$ &
$(\alpha_{\nu}, \beta_{\nu}, \sigma_{int})_{ml}$ &
$(\alpha_{\nu})_{-1\sigma \ -2\sigma}^{+1\sigma \ +2\sigma}$ &
$(\sigma_{int})_{-1\sigma \ -2\sigma}^{+1\sigma \ +2\sigma}$ \\
\hline \hline

~ & ~ & ~ & ~ & ~ & ~ & ~ \\

$u'$ & $(0.0, 0.680, 0.0)$ & $0.0_{-0.201 \ -0.416}^{+0.202 \ +0.416}$ &
$0.060_{-0.043 \ -0.057}^{+0.077 \ +0.211}$
& $(0.039, 0.146, 0.0)$ & $0.039_{-0.131 \ -0.192}^{+0.130 \ +0.270}$ &
$0.041_{-0.029 \ -0.039}^{+0.054 \ +0.145}$ \\

~ & ~ & ~ & ~ & ~ & ~ & ~ \\

$g'$ & $(0.037, -0.115, 0.0)$ & $0.037_{-0.135 \ -0.235}^{+0.176 \ +0.276}$ &
$0.043_{-0.030 \ -0.041}^{+0.054 \ +0.144}$
& $(0.050, 0.025, 0.0)$ & $0.050_{-0.093 \ -0.136}^{+0.139 \ +0.237}$ &
$0.031_{-0.022 \ -0.031}^{+0.040 \ +0.106}$ \\

~ & ~ & ~ & ~ & ~ & ~ & ~ \\

$r'$ & $(0.021, 0.115, 0.0)$ & $0.022_{-0.128 \ -0.262}^{+0.129 \ +0.264}$ &
$0.041_{-0.029 \ -0.039}^{+0.053 \ +0.140}$
& $(0.047, 0.075, 0.0)$ & $0.047_{-0.088 \ -0.179}^{+0.088 \ +0.180}$ &
$0.030_{-0.021 \ -0.047}^{+0.047 \ +0.099}$ \\

~ & ~ & ~ & ~ & ~ & ~ & ~ \\

$i'$ & $(0.004, 0.290, 0.0)$ & $0.005_{-0.112 \ -0.228}^{+0.112 \ +0.228}$ &
$0.039_{-0.028 \ -0.037}^{+0.048 \ +0.125}$
& $(0.052, 0.047, 0.0)$ & $0.052_{-0.077 \ -0.157}^{+0.078 \ +0.159}$ &
$0.029_{-0.021 \ -0.055}^{+0.055 \ +0.092}$ \\

~ & ~ & ~ & ~ & ~ & ~ & ~ \\

$z'$ & $(-0.012, 0.406, 0.0)$ & $-0.010_{-0.145 \ -0.285}^{+0.125 \ +0.265}$ &
$0.043_{-0.031 \ -0.041}^{+0.054 \ +0.143}$
& $(0.044, 0.114, 0.0)$ & $0.045_{-0.094 \ -0.191}^{+0.094 \ +0.192}$ &
$0.031_{-0.022 \ -0.031}^{+0.039 \ +0.103}$ \\

~ & ~ & ~ & ~ & ~ & ~ & ~ \\

\hline
\end{tabular}
\end{center}
\end{table*}

\subsection{Luminosity scaling relations}

Eqs.(\ref{eq: mlvsl}) and (\ref{eq: nvsl}) are linear relations (although in a logarithmic space) so that to determine their parameters we can resort to a general Bayesian procedure described in detail in D'Agostini (2005). Let us suppose that $(R, Q)$ are two quantities related by a power\,-\,law relation as $R = B Q^A$ with a certain intrinsic scatter $\sigma_{int}$. In logarithmic units, this reads $\log{R} = \alpha_l \log{Q} + \beta_l$ with $\alpha_l = A$ and $\beta_l = \log{B}$. In order to determine the parameters $(\alpha, \beta, \sigma_{int})$, we then maximize the following likelihood function ${\cal{L}} = \exp{[-L(\alpha_l, \beta_l, \sigma_{int})]}$ with\,:

\begin{eqnarray}
L(\alpha_l, \beta_l, \sigma_{int}) & = &
\frac{1}{2} \sum{\ln{(\sigma_{int}^2 + \sigma_{y_i}^2 + \alpha_l^2
\sigma_{x_i}^2)}} \nonumber \\
~ & + & \frac{1}{2} \sum{\frac{(y_i - \alpha_l x_i - \beta_l)^2}{\sigma_{int}^2 + \sigma_{y_i}^2 +
\alpha_l^2 \sigma_{x_i}^2}}
\label{eq: deflike}
\end{eqnarray}
where $(x_i, y_i) = (\log{Q_i}, \log{R_i})$ and the sum is over the ${\cal{N}}$ objects in the sample. Note that, actually, this maximization
is performed in the two\,-\,parameter space $(\alpha_l, \sigma_{int})$ since $\beta_l$ may be estimated analytically as\,:

\begin{displaymath}
\beta_l = \left [ \sum{\frac{y_i - \alpha_l x_i}{\sigma_{int}^2 + \sigma_{y_i}^2 + \alpha_l^2
\sigma_{x_i}^2}} \right ] \left [\sum{\frac{1}{\sigma_{int}^2 + \sigma_{y_i}^2 + \alpha_l^2
\sigma_{x_i}^2}} \right ]^{-1}
\label{eq: calca}
\end{displaymath}
so that we will no longer consider it as a fit parameter. The median values and confidence intervals for a given quantity can then be determined by studying the shape of the corresponding marginalized likelihood, i.e. the integral of ${\cal{L}}$ over the other parameter.

Let us then determine $\alpha_{\star}$ and $\alpha_{\nu}$ using the method outlined above. Note that these quantities depend on the wavelength adopted, but not on the anisotropy profile or MOND interpolating function. As a preliminary task, we bin the galaxies in roughly equally populated luminosity bins and remove all the galaxies with $n(f) \notin (2.0, 5.5)$ and problems with the absolute magnitude estimate. For a given luminosity bin and filter, we then analyse the distributions of $\Yst$ and $n$ and assign to that bin a value for $\Yst$ and $n$ using the median as central value and the $68\%$ confidence range as uncertainty. Note that, since such distributions may also be asymmetric, the errors on $\Yst$ and $n$ may turn out to be asymmetric. However, Eqs.(\ref{eq: deflike}) and (\ref{eq: calca}) assume that the uncertainties are symmetric. We therefore follow D'\,Agostini (2004) and correct the observed values as\,:

\begin{displaymath}
y_{corr} = y_{obs} + (\Delta_{+} - \Delta_{-}) \ \ ,
\ \ \sigma_{y} = (\Delta_{+} + \Delta_{-})/2 \ \ ,
\end{displaymath}
with $y_{obs}$ the median value, $(y_{min}, y_{max})$ its $68\%$ confidence range and $\Delta_{+} = y_{max} - y_{obs}$, $\Delta_{-} = y_{obs} - y_{min}$. It is worth stressing that such a correction is actually quite small so that we are confident that it is not biasing anyway the estimate of $(\alpha_{\star}, \alpha_{\kappa})$. The fit results are summarized in Table 1, where we report the values of the calibration parameters and the intrinsic scatter for the different filters.

As a general result, we find that both $\alpha_{\star}$ and $\alpha_{\nu}$ are quite small and, within the large error bars, are compatible with zero. Even if we consider only the best fit values, it is nevertheless clear that the correlation is quite shallow. However, the wide confidence ranges on $(\alpha_{\star}, \alpha_{\nu})$ prevent us from drawing any definitive conclusion other than the qualitative observation that $\Yst$ and $n$ only weakly correlate with the luminosity. While the result for the \Yst\,-\,$L$ correlation is in agreement with previous results in the literature, the very shallow slope of the $n$\,-\,$L$ relation is somewhat at odds with previous findings. For instance, using galaxies with well determined photometric parameters in the Virgo sample, Nipoti et al. (2008) found $n \propto L_B^{0.27 \pm 0.02}$ which is definitively larger than our best fit $\alpha_{\nu}$ for the $g'$ filter (the closest to the $B$ one), although well within the wide $68\%$ confidence range. Such a discrepancy is actually illusory and is due to the combination of the small luminosity range probed and the large error bars. Indeed, our ETGs sample actually spans approximately just one order of magnitude in luminosity (the first bin being at 9.1 and the last at 10.4). Adopting the Nipoti et al. slope, this leads to a change in $n$ of order 0.35 which is lower than the width of the $68\%$ confidence range for $n$ in a given bin. Indeed, if we fix the slope of the $n$\,-\,$L$ relation in the $g'$ filter to the one given by Nipoti et al. and only adjust the zeropoint and the intrinsic scatter, we get a very satisfactory result. We therefore argue that no discrepancy is actually present. As an alternative, we could have fitted the $n$\,-\,$L$ relation without dividing galaxies into luminosity bins. However, binning allows to wash out the biases in the $n$ determination which are inherited by the SDSS photometric fitting code so that we have preferred to use this procedure even at the price of having larger error bars.

\subsection{Predicting the MFP coefficients}

The $(\alpha_{MFP}, \beta_{MFP})$ values may be estimated using Eq.(\ref{eq: mfpcoeff}) provided one has set the filter (and hence the corresponding value of $\alpha_{\star}$ and $\alpha_{\nu}$) and evaluated the $(a, b, c)$ coefficients. However, we do not have a single value for all the quantities of interest, but rather a distribution that, as a first order approximation, we can model as a Gaussian one with central value and standard deviation obtained by correcting the asymmetries in the inferred distributions. In order to estimate $(\alpha_{MFP}, \beta_{MFP})$, for a given combination of filter, anisotropy profile and MOND interpolating function, we generate $(\beta, \alpha_{\star}, \alpha_{\nu})$ from a multinormal distribution and use Eq.(\ref{eq: mfpcoeff}) to compute the MFP parameters. We repeat this procedure 10000 times and use the distribution thus obtained to estimate the mean and median values and the $68$ and $95\%$ confidence ranges. To this end, we use the $\beta$ distribution estimated as described in Sect. 5.4.3 since that procedure allows us to both get physical consistent models and minimize the deviations from the fiducial $\Yst$ value.

\begin{figure}
\includegraphics[width=8cm]{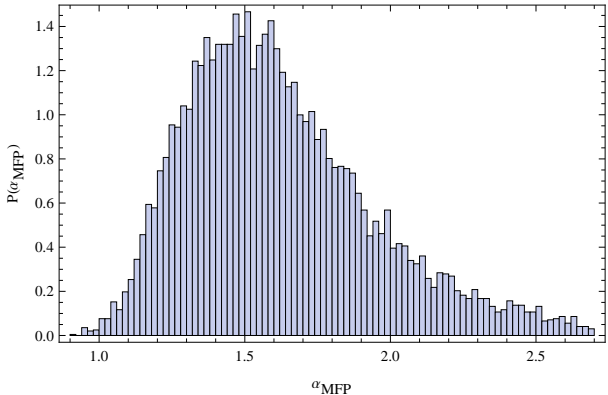}
\caption{Histogram of the $\alpha_{MFP}$ values obtained in the $i'$ filter.}
\label{fig: alphasimple}
\end{figure}

\begin{figure}
\includegraphics[width=8cm]{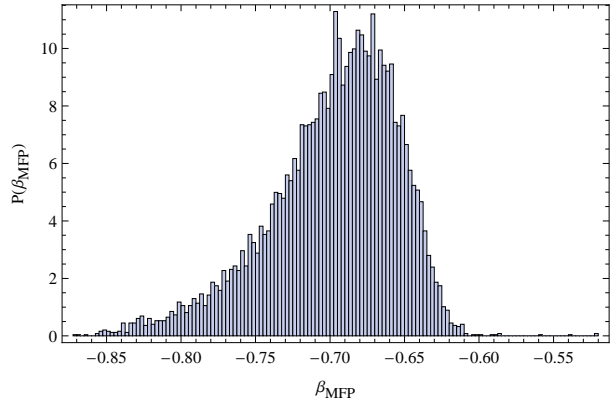}
\caption{Same as Fig.\,\ref{fig: alphasimple} but for the $\beta_{MFP}$ parameter.}
\label{fig: betasimple}
\end{figure}

The resulting $\alpha_{MFP}$ and $\beta_{MFP}$ distributions for the $i'$ filter are shown in Figs.\,\ref{fig: alphasimple} and \ref{fig: betasimple} and can be quantitatively summarized by the mean and median values with confidence ranges given below\,:

\begin{displaymath}
\langle \alpha_{MFP} \rangle = 1.62 \ \ , \ \
(\alpha_{MFP})_{med} = 1.58 \ \ ,
\end{displaymath}
\begin{displaymath}
{\rm 68\% \ CL \ :} \ \ (1.31, 1.94) \ \ , \ \
{\rm 95\% \ CL \ :} \ \ (1.13, 2.42) \ \ ,
\end{displaymath}

\begin{displaymath}
\langle \beta_{MFP} \rangle = -0.70 \ \ , \ \
(\beta_{MFP})_{med} = -0.69 \ \ ,
\end{displaymath}
\begin{displaymath}
{\rm 68\% \ CL \ :} \ \ (-0.74, -0.66) \ \ , \ \
{\rm 95\% \ CL \ :} \ \ (-0.80, -0.63) \ \ .
\end{displaymath}
As it is apparent from Figs.\,\ref{fig: alphasimple} and \ref{fig: betasimple}, while the $\alpha_{MFP}$ distribution is quite wide, the $\beta_{MFP}$ one is, on the contrary, quite narrow thus leading to small errors on this quantity. Such a result can be explained by noting that, since $(\alpha_{\star}, \alpha_{\nu})$ are quite small, Eq.(\ref{eq: mfpcoeff}) approximately reduce to $\alpha_{MFP} \simeq a$ and $\beta_{MFP} \simeq b$. Since $a$ depends more strongly on $\beta$ than on $b$, the scatter of $\beta$ around the median value has a stronger impact on $\alpha_{MFP}$ than on $\beta_{MFP}$ thus explaining why the distribution of $\alpha_{MFP}$ is wider than the distribution of $\beta_{MFP}$.

Both the mean values and the $68\%$ confidence ranges are almost unchanged if one replaces the simple MOND interpolating function with the standard one. This result is encouraging since it tells us that the arbitrariness in the choice of the simple or standard function does not bias the final MFP coefficients. Actually, such a conclusion should be better tested by considering other $\mu(g/a_0)$ expressions leading to different $\gamma_{MOND}(r)$ to be inserted in Eq.(\ref{eq: sigmalosgen}). We can, however, argue that the effect will be quite weak as can be easily explained by looking at the two $\gamma_{MOND}(\eta)$ functions in Eqs.(\ref{eq: simplegamma}) and (\ref{eq: standgamma}). Both expressions may be roughly approximated by $\gamma_{MOND} \simeq 1 + (a_0/a_{eff}) f(\eta)$ with the details of the function $f(\eta)$ depending on the adopted expression of $\mu(g/a_0)$. For typical values of the ETGs parameters, $a_0/a_{eff} << 1$ so that the details of the $f(\eta)$ function are important only for very large values of $\eta$. However, the contribution to $\sigma_0$ mainly comes from the inner regions thus explaining why the choice of the MOND interpolating function only negligibly affects the estimate of the MFP coefficients.

In the Newtonian framework, an easy application of the virial theorem (and the hypotheses of homology and constant $M/L$ ratio) leads to what we have referred to as the virial plane, namely a FP\,-\,like relation with coefficients $(\alpha_{vir}, \beta_{vir}) = (2, -1)$. The MFP coefficients
are definitely different from the virial ones. This is only partly due to the use of the scaling relations (\ref{eq: mlvsl}) and (\ref{eq: nvsl}). If we turn off their effect by forcing $\alpha_{\star} = \alpha_{\nu} = 0$, we should have obtained values for $(\alpha_{MFP}, \beta_{MFP})$ quite similar to those reported here. It is actually the combined effect of the anisotropy and MOND that leads to this departure from the Newtonian virial predictions. The effect is however quite weak on $\alpha_{MFP}$ which only mildly departs from $\alpha_{vir} = 2$ remaining consistent with this value within the errors, while it is stronger for $\beta_{MFP}$ with the virial value $\beta_{vir} = -1$ definitely out of the $95\%$ confidence range. We therefore argue that MOND is indeed able to introduce departure from the Newtonian virial plane even if the ETGs inner regions are for the most part in the Newtonian regime.

The disagreement between the virial FP coefficients and the observed ones is usually referred to as the problem of the FP tilt. The introduction of luminosity scaling relations for $n$ and $\Yst$ is unable to explain the tilt so that one postulates the presence of a dark matter halo representing $\sim 30 - 50\%$ of the mass within $R_{eff}$ and providing a total $M/L$ ratio scaling with luminosity as $\Upsilon_{tot} \propto L^{\alpha_{\Upsilon}}$ with $\alpha_{\Upsilon} \simeq 0.15 - 0.25$. It is worth wondering whether the FP tilt may instead be explained by our MOND based models. Should this be the case, the MFP coefficients should turn out to be consistent with the observed FP coefficients. Actually, there are different estimates of the FP coefficients relying on different samples and different fitting procedures so that a fair comparison is not easy (see, e.g., Bernardi et al. (2003) for a table with some estimates). We have therefore used our Bayesian method (generalizing from 2 to 3 parameters the previous formulae) to find as best fit parameters\,:

\begin{displaymath}
(\alpha_{obs}, \beta_{obs}, \sigma_{int}) = (1.13, -0.75, 0.081)
\end{displaymath}
in good agreement with $(\alpha_{obs}, \beta_{obs}, \sigma_{int}) = (1.14 \pm 0.04, -0.76 \pm 0.01, 0.085)$ in Bernardi et al. (2003) which use a direct fit method and correct for evolution and selection effects. The median value and the $68\%$ and $95\%$ confidence ranges read\footnote{Note that, to save computer time, we have fitted not the full sample, but a subsample made out by 1000 randomly selected ETGs. While this has no effect on the best fit parameters (as we have explicitly checked), it is likely that the $95\%$ confidence range is wider than what we would have obtained using the full sample.}\,:

\begin{displaymath}
\alpha_{obs} = 0.96_{-0.30 \ -0.52}^{+0.33 \ +0.69} \ \ , \ \
\beta_{obs} = -0.63_{-0.21 \ -0.34}^{+0.14 \ +0.25} \ \ .
\end{displaymath}
The comparison of the predicted MFP coefficients with the observed ones shows that the tilt of the FP has been only partially reduced. Indeed, if we consider the best fit values for $(\alpha_{obs}, \beta_{obs})$, we find that, while $\beta_{obs}$ is nicely close to the median $\beta_{MFP}$, $\alpha_{obs}$ is definitely smaller than the median $\alpha_{MFP}$. However, in a Bayesian framework, what is more important is the comparison with the marginalized confidence ranges. In this case, $\beta_{MFP}$ and $\beta_{obs}$ are in remarkable good agreement, while the $95\%$ confidence ranges of $\alpha_{MFP}$ and $\alpha_{obs}$ have only a small overlap. We therefore conclude that there is only a mild chance that a MOND based approach align the observed FP with a theoretically predicted one.

An important remark is in order here. We have shown, in Sect. 5.4, that MOND is consistent with the observed ETGs dynamics, that is to say, given $(n, R_{eff}, \langle I_e \rangle)$ and the stellar $M/L$ ratio \Yst, we can find a value for $\beta$ such that the theoretically $\sigma_{ap}$ equals the observed one. We have checked that this condition is enough to guarantee that the same matching still works for $\sigma_0$ with $\sigma_0$ estimated from $\sigma_{ap}$ through Eq.(\ref{eq: sigmacorr}). Since a galaxy is placed on the observed FP according to its $\sigma_0$ value, predicting the correct $\sigma_0$ enables us to correctly place the galaxy on the observed FP. Nevertheless, our theoretical MFP is tilted with respect to the observed one. This inconsistency originates from the following reason. When one considers an individual galaxy, the value of $\beta$ is set according to the values of $(n, R_{eff}, \langle I_e \rangle, \Yst)$ of that particular galaxy. On the contrary, the theoretical MFP is derived from a model that is assumed to be the same for all the galaxies. In particular, the value of $\beta$ used to get $(\alpha_{MFP}, \beta_{MFP})$ is the same for all the galaxies with the error on the MFP coefficients deriving from the width of the $\beta$ distribution. Of course, $\beta$ is not the same for all the galaxies so that one can not use the theoretical MFP to predict the $\sigma_0$ value for a given galaxy. In other words, one could always find $(\alpha_{MFP}, \beta_{MFP})$ for each individual galaxy; however, the set of these values does not define a single MFP, but rather a set of ${\cal{N}}$ MFPs different from each other. The fact that this collection of planes does not reduce to a single one is another way of saying that the observed FP and the theoretical MFP are tilted.

\section{Conclusions}

The MOND framework has nowadays a long history of successes when applied to galaxy scale systems: it was shown to be able to efficiently explain spiral galaxy flat rotation curves, and empirical scaling relations such as the Tully\,-\,Fisher law and the Baryonic Tully\,-\,Fisher relation (see Sanders \& McGaugh 2002 and refs. therein). In the standard Newtonian scenario, dark matter is invoked not only in spiral galaxies, but also in ETGs to both reproduce the observed dynamics and explain the FP tilt. Since MOND is, by construction, a universal theory, one should be able to remove dark matter in ETGs as in spiral galaxies. Motivated by this consideration, we have investigated whether this is the case and derived an
FP\,-\,like relation which we have referred to as the MOND fundamental plane (MFP).

To this aim, we have extracted from the NYU\,-\,VAGC lowZ catalog a sample of $\simeq 9000$ ETGs with accurate five bands photometry and measured aperture velocity dispersion. Assuming that no dark matter is present, we can simply model an elliptical galaxy using the Prugniel and Simien (1987) model, setting the $(n, R_{eff}, \langle I_e \rangle)$ parameters from photometry and inferring the stellar $M/L$ ratio $\Yst$ from the observed colours. We then consider a constant velocity anisotropy profile and the simple form for the MOND interpolating function $\mu(g/a_0)$ and constrain the anisotropy parameter $\beta$ by matching the predicted and the observed aperture velocity dispersions. As a first important result, we find that physically consistent MOND models can be found to fit the data provided a radial velocity anisotropy is assumed and the dynamical $M/L$ ratio is adjusted within the uncertainties of the stellar $M/L$. These results are independent of the choice of the (poorly constrained) MOND interpolating function $\mu(x)$ so that we may safely argue that MOND can remove dark matter not only in spiral galaxies, but also in early\,-\,type ones.

Fitting the velocity dispersion data also allows us to infer the distribution of the anisotropy parameter and hence estimate the MFP slope coefficients $(\alpha_{MFP}, \beta_{MFP})$. We find that these quantities are different from those expected when applying the virial theorem (with homology and constant $M/L$ ratio assumptions) in the Newtonian framework, i.e. MOND is able to tilt the virial plane without the need of assuming a varying dark matter content. Such a tilt is, however, still too small to align the MFP with the observed FP. Indeed, we find that, while $\beta_{MFP}$ is in remarkable good agreement with $\beta_{obs}$, $\alpha_{MFP}$ is significantly larger than $\alpha_{obs}$ so that we need a mechanism to further tilt the MFP.

In particular, one could explore deviations from the pure MOND scenario we have used throughout the paper. Two possibilities are briefly hinted at here and discussed in some detail in Appendix B. As a first scenario, a varying $M/L$ ratio can reconcile the predicted MFP with the observed FP provided that $\kappa_{eff} = \Upsilon_{eff}/\Yst \propto L^{0.26}$. A similar scaling might be expected in the MOND model by Angus (2009) with 11~eV sterile neutrinos. Angus et al. (2010) show that, in this model, more massive galaxy clusters, that generally harbour the most massive and luminous central galaxies, have larger central densities of sterile neutrinos and, presumably, a larger total $M/L$ ratio.

As a totally different approach, one can rely on the external field effect (EFE). Different from the varying $M/L$ scenario, the EFE is naturally motivated because it is a consequence of the nonlocal feature of the MOND theory. The inclusion of EFE is able to tilt the MFP without the need of additional matter and without resorting to some fine tuned mechanism. The scaling of $g_{ext}/a_0$ with $L$ needed to reconcile the theoretical MFP with the observed FP is reasonable, but has to be verified by a careful analysis. To this end, an ideal approach could be fitting the full velocity dispersion profile $\sigma_{los}(R)$ rather than the aperture value $\sigma_{ap}$. This test has to be carried on for a statistically meaningful sample of ETGs with $\sigma_{los}(R)$ measured both in the inner regions (where the EFE has a small impact) and in the outskirts (where the anisotropy profile reduces to its asymptotic value) so that the $\beta$\,-\,$g_{ext}$ degeneracy is broken. Moreover, this sample should cover a wide range in luminosity and probe different environments.

As a final comment, we would like to stress that, although the two scenarios proposed above are, at the moment, only speculative, reconciling the theoretical MFP with the observed FP and reproducing other ETGs scaling relations in the MOND framework can open a new and fruitful way towards discriminating between modified dynamics (or, more generally, modified gravity theories) and dark matter.

\section*{Acknowledgements}

We thank the referee for his insightful comments that helped us to improve the paper. VFC and GWA are supported by Regione Piemonte and Universit\`a di Torino. Partial support to VFC, GWA and AD from INFN project PD51 and the PRIN-MIUR-2008 grant ``Matter-antimatter asymmetry, dark matter
and dark energy in the LHC Era'' is also acknowledged. CT is supported by the Swiss National Science Foundation and  by a grant from the project Mecenass, funded by Compagnia di San Paolo.

\appendix

\section{Wavelength dependence}

Eq.(\ref{eq: mfpcoeff}) shows that the MFP coefficients depend on the slopes $(\alpha_{\star}, \alpha_{\nu})$ of the $\Yst$\,-\,$L_T$ and $n$\,-\,$L_T$ relations which are different depending on the filter considered, as reported in Table 1. Moreover, as explained in Sect.\,5.4, the distribution of the anisotropy parameter also depends on the filter adopted because $(n, R_{eff}, \langle I_e \rangle, \Yst)$ vary with the filter. As a consequence of these effects, the MFP coefficients will depend on wavelength. To check how they change with $\lambda$, we report in Table A1 the values of $(\alpha_{MFP}, \beta_{MFP})$ for the fiducial case of constant anisotropy with the simple MOND interpolating function.

\begin{table}
\caption{The MFP coefficients for the fiducial case in different filters. Columns are as in Table 2, while the
rows are for filters $u'$, $g'$, $r'$, $i'$, $z'$ respectively.}
\begin{center}
\begin{tabular}{|c|c|c|c|}
\hline
$\langle \alpha_{MFP} \rangle$ &
$(\alpha_{MFP})_{-1\sigma \ -2\sigma}^{+1\sigma \ +2\sigma}$ &
$\langle \beta_{MFP} \rangle$ &
$(\beta_{MFP})_{-1\sigma \ -2\sigma}^{+1\sigma \ +2\sigma}$ \\
\hline \hline

~ & ~ & ~ & ~ \\

1.67 & $1.58_{-0.37 \ -0.61}^{+0.60 \ +1.21}$ & -0.70 & $-0.69_{-0.07 \ -0.15}^{+0.05 \ +0.08}$ \\

~ & ~ & ~ & ~ \\

1.44 & $1.36_{-0.27 \ -0.46}^{+0.45 \ +1.12}$ & -0.68 & $-0.67_{-0.05 \ -0.13}^{+0.04 \ +0.06}$ \\

~ & ~ & ~ & ~ \\

1.67 & $1.60_{-0.30 \ -0.50}^{+0.45 \ +1.02}$ & -0.71 & $-0.70_{-0.06 \ -0.13}^{+0.04 \ +0.07}$ \\

~ & ~ & ~ & ~ \\

1.62 & $1.58_{-0.27 \ -0.45}^{+0.36 \ +0.84}$ & -0.70 & $-0.69_{-0.05 \ -0.11}^{+0.03 \ +0.06}$ \\

~ & ~ & ~ & ~ \\

1.67 & $1.60_{-0.31 \ -0.52}^{+0.47 \ +1.07}$ & -0.71 & $-0.60_{-0.06 \ -0.14}^{+0.04 \ +0.07}$ \\

~ & ~ & ~ & ~ \\

\hline
\end{tabular}
\end{center}
\end{table}
It is immediate to see that the MFP coefficients are essentially the same (within the errors) whatever is the adopted filter. This is an expected result since both $(\alpha_{\star}, \alpha_{\nu})$ and the $\beta$ distribution are very weak function of wavelength. This is in net contradiction with the observed FP coefficients that move towards the virial one $(2, -1)$ as the wavelength increases. As a consequence, the tilt of the MFP changes with the filter thus leading further constraints on the underlying unknown phenomenon that has to be invoked to reconcile theory with observations.

\section{Deviations from a pure MOND scenario}

Up to now, we have adopted a fully MOND framework. Specifically, we have assumed that the visible matter is the only source of gravity generating the motion of the stars and that their acceleration may be computed from the Newtonian one with Eq.(\ref{eq: mondgeq}). As we have seen, this model is able to fit the ETG velocity dispersion data, but leads to an MFP which is still tilted with respect to the observed one. We therefore qualitatively discuss here what could be the impact of relaxing one of the two assumptions above to see whether deviations from our idealized MOND scenario may help to reconcile the theoretical MFP with the observed FP. We admit that this section is quite speculative, but we include it here in order to propose some scenarios that could be tested with future data or a different tracer of the ETGs gravitational potential.

\subsection{Varying $M/L$ ratio}

According to the original idea motivating the introduction of MOND, nothing but the visible matter should be considered when modelling a galaxy and trying to fit the data. As a corollary to this assumption, the total $M/L$ ratio $\Upsilon_{tot}$ must equal the  stellar $\Yst$. Even if in order to avoid too much radial anisotropy we have allowed to rescale the $M/L$ by the parameter $\kappa$, we have finally chosen the solution with $\kappa$ as close as possible to $1$ to be consistent with the idea of no dark matter.  Some general considerations, however, may be invoked to depart from the $\Upsilon_{tot} \simeq \Yst$ ansatz. First, it is well known that MOND needs a dark matter component in order to be in agreement with data on the galaxy cluster scale (Aguirre et al. 2001; Sanders 2003; Pointconteau \& Silk 2005; Angus et al. 2007, 2008a). Should this further term be represented by a 2\,eV neutrinos (Sanders 2003, 2007) or 11\,eV sterile neutrinos (Angus 2009; Angus et al. 2010) arranged in a hot dark halo, the total $M/L$ ratio should be larger than the stellar one. On the other hand, our estimated $\Yst$ depends on the ingredients used to obtain the starting library of stellar population models. Changing one of these ingredients would alter $\Yst$ possibly leading to an effective $\Upsilon_{tot}$ larger than our assumed stellar $M/L$ value. As a simple way to account for this possibility, we redefine the effective surface mass density as\,:

\begin{equation}
\langle \Sigma_e \rangle = \Upsilon_{tot} \langle I_e \rangle =
\kappa_{\Sigma} \Yst \langle I_e \rangle \ .
\label{eq: kappadef}
\end{equation}
If we assume that $\kappa_{\Sigma}$ correlates with luminosity and that the $\log{\kappa_{\Sigma}}$\,-\,$\log{L}$ may be well approximated by\,:

\begin{equation}
\log{\kappa_{\Sigma}} = \alpha_{\kappa} \log{L} + \beta_{\kappa} \ ,
\label{eq: kappavsl}
\end{equation}
we can easily rederive the MFP to show that Eq.(\ref{eq: mfp}) still holds provided the coefficients are redefined as\,:

\begin{equation}
\left \{
\begin{array}{l}
\displaystyle{\alpha_{MFP} = \frac{a}{1 - 2 \alpha_{\Upsilon} b - 2 \alpha_{\nu} c}} \\
~ \\
\displaystyle{\beta_{MFP} = \frac{(1 + \alpha_{\Upsilon}) b + \alpha_{\nu} c}
{1 - 2 \alpha_{\Upsilon} b - 2 \alpha_{\nu} c}} \\
~ \\
\displaystyle{\gamma_{MFP} = \frac{(\alpha_{\Upsilon} b + \alpha_{\nu} c)
\log{(2 \pi)} + \beta_{\Upsilon} + b + \beta_{\nu} c + d}
{1 - 2 \alpha_{\Upsilon} b - 2 \alpha_{\nu} c}} \\
\end{array}
\right .
\label{eq: mfpcoeffbis}
\end{equation}
with

\begin{equation}
\left \{
\begin{array}{l}
\displaystyle{\alpha_{\Upsilon} = \alpha_{\star} + \alpha_{\kappa}} \\
~ \\
\displaystyle{\beta_{\Upsilon} = \beta_{\star} + \beta_{\kappa}} \\
\end{array}
\right . \ .
\label{eq: defalphatot}
\end{equation}
Needless to say, introducing $\kappa_{\Sigma}$ changes the distribution of the fitted anisotropy parameter and the values of
the $(a, b, c, d)$ coefficients. Assuming, as a first approximation, that the change in $\beta$ is not too large, we solve $\alpha_{MFP} = \alpha_{obs}$ with respect to $\alpha_{\kappa}$. By using the median values of $(\beta, \alpha_{\star}, \alpha_{\nu})$ for the constant anisotropy model and $\alpha_{obs} = 1.13$, we find $\alpha_{\kappa} \simeq 0.35$. It is worth noting that a similar scaling is obtained when comparing stellar and dynamical masses in a Newtonian stellar+dark halo framework \cite{Pad04,cresc08a,CTMS09}. Actually, reproducing such a scaling of $\kappa_{\Sigma}$ with luminosity can be problematic in a MOND scenario. As quoted above, in order to have $\kappa_{\Sigma} \neq 1$, one has to postulate the presence of a halo made of (2eV or 11eV) neutrinos. We now find that such a term should provide a contribution to the velocity dispersion that increases with the total luminosity in a similar way as the dark halo in Newtonian gravity. Explaining how neutrinos in the halo interact with the baryons in the stellar component in such a way to reproduce the needed scaling of $\kappa_{\Sigma}$ with $L$ can be a difficult hurdle to overcome.

Should such a mechanism be found, the problem of the tilt of the MFP is indeed solved. Introducing $\alpha_{\kappa} = 0.35$ in the expression for $\beta_{MFP}$ and using the median values of $(\beta, \alpha_{\star}, \alpha_{\nu})$ gives $\beta_{MFP} \simeq -0.64$ which is essentially the same as for the $\kappa_{\Sigma} = 1$ case. This can be explained by looking at the approximated expression reported above showing that the same term $\alpha_\Upsilon b$ enters both the numerator and the denominator thus weakening the correction. As a result, the predicted $\beta_{MFP}$ is still in agreement with the observed one thus completely solving the tilt problem.

\subsection{The external field effect}

As a second possibility, we consider modifying the relation between Newtonian and MOND acceleration. We note that Eq.(\ref{eq: mondgeq}) implicitly assumes that the galaxy is an isolated system. Since MOND is a non-local theory, the motion of the stars in the galaxy is actually determined not only by the galaxy potential, but also by the distribution of matter outside the galaxy itself. This is usually referred to as the {\it external field effect} (EFE) and is taken into account by replacing Eq.(\ref{eq: mondgeq}) with the following one\,:

\begin{equation}
\mu\left ( \frac{g + g_{ext}}{a_0} \right ) g = g_N
\label{eq: mondgeqefe}
\end{equation}
where $g_{ext}$ is a constant (with typical values $\sim 0.1 - 10 \ a_0$). In order to take the EFE into account, we must simply use the corresponding $\gamma_{MOND}(\eta)$ function which now reads\,:

\begin{eqnarray}
\gamma_{MOND}(\eta) & = & \frac{1}{2} \left \{ \left ( 1 - \frac{g_{ext}}{g_{N}(\eta)} \right ) \right . \nonumber \\
~ & + & \left . \sqrt{1 + \frac{4 a_0}{g_N(\eta)} + \frac{2 g_{ext}}{g_N(\eta)} + \left [ \frac{g_{ext}}{g_N(\eta)} \right ]^2} \right \}
\label{eq: efegamma}
\end{eqnarray}
with $g_N(\eta) = G M(\eta)/\eta^2$. Comparing Eq.(\ref{eq: efegamma}) with (\ref{eq: simplegamma}) shows that, for given stellar parameters,
$\gamma_{MOND}(\eta)$ for EFE is smaller than the one for the simple function so that the predicted velocity dispersion is smaller. Such a
somewhat counterintuitive result can be qualitatively explained by noting that the EFE term $g_{ext}$ increases the Newtonian acceleration and delays the transition from the Newtonian to the MOND regime. As a consequence, the boost in acceleration (and hence in $\sigma_0$) due to the MOND effect starts later and reduces $\sigma_0$ with respect to the no EFE  case.

In order to study the impact of the EFE on the MFP coefficients, we start from the model with constant anisotropy and no EFE and rewrite Eq.(\ref{eq: premfp}) as\,:

\begin{displaymath}
\log{R_{eff}} = a \log{\left ( \frac{\sigma_0}{\sigma_0^{EFE}} \times \sigma_0^{EFE} \right )}
+ b \log{\langle \Sigma_e \rangle} + c \log{n} + d
\end{displaymath}
with $\sigma_0$ and $\sigma_0^{EFE}$ the velocity dispersion without and with the EFE taken into account. Note that proceeding this way allows
us to estimate $(a, b, c, d)$ using the values found for the no EFE simple case. In a first reasonably good approximation, it is\,:

\begin{displaymath}
\log{\left ( \frac{\sigma_0}{\sigma_0^{EFE}} \right )} \simeq a_{ext} \log{\left ( \frac{g_{ext}}{a_0} \right )} + b_{ext}
\end{displaymath}
with $(a_{ext}, b_{ext}) \simeq (0.09, 0.21)$. If we further postulate that $g_{ext}$ correlate with luminosity as

\begin{displaymath}
\log{(g_{ext}/a_0)} = \alpha_{ext} \log{L} + \beta_{ext}
\end{displaymath}
and repeat the same steps leading from Eq.(\ref{eq: premfp}) to Eq.(\ref{eq: mfp}), we find that the slope coefficients of the MFP now read\,:

\begin{equation}
\left \{
\begin{array}{l}
\displaystyle{\alpha_{MFP} = \frac{a}{1 - 2 ( \alpha_{\star} b + \alpha_{\nu} c + a a_{ext} \alpha_{ext})}} \\
~ \\
\displaystyle{\beta_{MFP} = \frac{(1 + \alpha_{\star}) b + \alpha_{\nu} c + a a_{ext} \alpha_{ext}}
{1 - 2 ( \alpha_{\star} b + \alpha_{\nu} c + a a_{ext} \alpha_{ext})}} \\
\end{array}
\right .
\label{eq: mfpcoeffefe}
\end{equation}
which reduce to Eq.(\ref{eq: mfpcoeff}) for $a_{ext} = 0$, i.e. no EFE effect. In order to get an estimate of $\alpha_{ext}$, we can assume that the values of $(a, b, c)$ are the same as those computed using the median $\beta$ value obtained by fitting the constant anisotropy model with no EFE to the data. Setting $(\alpha_{\star}, \alpha_{\nu})$ to the median values in Table 1 for the $i'$ filter and solving $\alpha_{MFP} = \alpha_{obs}$, we finally get $\alpha_{ext} \simeq -1.5$, i.e. brighter galaxies should be affected by a lower EFE.

Investigating whether such a correlation is observationally motivated is a rather difficult task. In order to estimate $g_{ext}$, one can not resort to a fit to galaxies binned in luminosity since the magnitude of the EFE depend on the environment in which a given galaxy is embedded. Although it is possible to study the environment of the ETGs in our sample on a case\,-\,by\,-\,case basis, their large number makes this task quite beyond the scope of this paper. As a general remark, we however note that the EFE is typically invoked to improve the fit to the rotation curves of dwarf spiral galaxies (see, e.g., Angus 2008), while the value of $g_{ext}$ is smaller for Milky Way like systems. This observation goes in the right direction. Moreover, one has to take care also of the symmetry of the problem. Brighter galaxies are typically more massive and hence populate the inner regions of clusters where they feel the combined action of many galaxies around. One can naively expect that the random orientation of the EFE from each companion leads to a sort of compensation thus lowering $g_{ext}$ for brighter systems. However, a further quantitative analysis is mandatory in order to lend support to this scenario.

\end{document}